\documentclass[lettersize,journal]{IEEEtran}
\usepackage{amsmath,amsfonts,amssymb}
\usepackage{colortbl}
\usepackage[ruled,linesnumbered]{algorithm2e}
\usepackage{array}
\usepackage{booktabs}
\usepackage{subfigure}
\usepackage{textcomp}
\usepackage{stfloats}
\usepackage{url}
\usepackage[normalem]{ulem} 
\usepackage{verbatim}
\usepackage{multicol}
\usepackage{multirow}
\usepackage{makecell}
\usepackage{graphicx}
\usepackage{hyperref}
\usepackage{forest}
\def\BibTeX{{\rm B\kern-.05em{\sc i\kern-.025em b}\kern-.08em
    T\kern-.1667em\lower.7ex\hbox{E}\kern-.125emX}}
\usepackage{balance}
\usepackage{tikz}
\usetikzlibrary{positioning}
\usetikzlibrary{backgrounds, fit, arrows.meta}
\usepackage{tikz-qtree}

\newtheorem{theorem}{Theorem}
\newtheorem{definition}{Definition}
\newtheorem{proposition}{Proposition}
\newtheorem{lemma}{Lemma}

\newtheorem{example}{Example}
\newtheorem{remark}{Remark}

\begin{document}
\title{A Parallelization Strategy for GRAND with Optimality Guarantee by Exploiting Error Pattern Tree Representation}

\author{Li Wan, Huarui Yin, Wenyi Zhang,~\IEEEmembership{Senior Member,~IEEE}
                   
\thanks{The authors are with Department of Electronic Engineering and Information Science, University of Science and Technology of China, Hefei, China. Emails: wan\_li@mail.ustc.edu.cn, \{yhr, wenyizha\}@ustc.edu.cn. This work was supported in part by the National Natural Science Foundation of China under Grant 62231022.}}


\maketitle

\begin{abstract}
    Parallelism has become a central concern in modern decoding frameworks aiming to meet stringent throughput and latency requirements. Guessing Random Additive Noise Decoding (GRAND) is a recently proposed decoding paradigm that tests candidate Error Patterns (EPs) until a valid codeword is found. Among its variants, Soft GRAND (SGRAND) achieves maximum-likelihood (ML) decoding but relies on real-time generation and likelihood ordering of EPs, making parallel execution nontrivial under the ML optimality constraint. In this work, we introduce a unified binary tree representation of EPs, termed the EP tree, which formalizes the hierarchical structure underlying SGRAND and Ordered Reliability Bits (ORB) GRAND algorithms, enabling structured organization of EPs and algorithmic-level parallel exploration. Building upon this unified framework, we propose a parallel design of SGRAND that preserves ML optimality while significantly reducing decoding complexity through pruning strategies and tree-based computation. Furthermore, we develop an enhanced ORBGRAND algorithm based on the same EP tree representation, improving decoding performance toward ML while retaining parallel efficiency. Numerical experiments show that the proposed parallel SGRAND achieves a $3.96\times$ reduction in decoding latency compared with its serial counterpart, while the enhanced ORBGRAND achieves a $4.21\times$ speedup, demonstrating the effectiveness of the unified tree-based framework and its strong potential for future algorithmic and hardware optimizations.
\end{abstract}

\section{Introduction}\label{sec:intro}

Advances in modern hardware platforms have made parallel execution an essential consideration in the design of practical channel decoders \cite{shao2019survey,leroux2012semi,tarver2021gpu}. However, for certain decoding paradigms, effectively exploiting parallelism while preserving desirable decoding behavior remains a nontrivial algorithmic challenge.

Guessing Random Additive Noise Decoding (GRAND) is a recently proposed decoding paradigm \cite{duffy2019capacity} with strong potential for universal and efficient algorithmic realization \cite{abbas2023guessing}. GRAND proceeds by systematically generating and testing error patterns (EPs) until a valid codeword is found. Its attractive performance at short-to-moderate block lengths and high code rates \cite{duffy2022ordered} makes it particularly well suited for ultra-reliable low-latency communication (URLLC) applications, such as autonomous driving and virtual/augmented reality, where stringent reliability and latency targets must be simultaneously satisfied \cite{shirvanimoghaddam2018short,wang2023road}. It should be noted that an upper bound on the decoding complexity of GRAND grows with the number of parity bits, and therefore GRAND remains highly challenging in regimes with a large number of parity bits \cite{yue2023efficient}.

For channels with soft output, two principal variants of GRAND have been proposed, each exhibiting complementary trade-offs between decoding optimality and algorithmic simplicity. Soft GRAND (SGRAND)~\cite{solomon2020soft} achieves maximum-likelihood (ML) decoding by ordering EPs according to their likelihoods derived from channel soft information. When multiple candidate EPs are generated, parallel processing can be realized by examining them simultaneously; however, how to obtain these EPs in parallel, and how to preserve the ML guarantee within a parallel framework have not been fully studied.
In contrast, ordered reliability bit (ORB) GRAND~\cite{duffy2022ordered} and its ORB-type variants~\cite{liu2022orbgrand,wan2024approaching} adopt an alternative EP ordering principle based solely on the ranking of channel reliabilities, rather than their exact likelihood values. This ranking-based formulation simplifies EP generation and naturally admits efficient parallel execution, at the cost of sacrificing strict ML optimality.

To reconcile these complementary properties, this paper introduces a unified, explicitly tree-based framework that captures the structural organization of EPs underlying both SGRAND and ORBGRAND. Based on this framework, we develop a parallel decoding algorithm for SGRAND and design an enhanced version of ORBGRAND capable of achieving ML decoding. The key idea is to formalize the underlying parent–child relationships among EPs (as discussed in \cite{solomon2020soft}) into an explicit binary-tree representation termed the EP tree. This representation provides a unified view that supports concurrent likelihood evaluation, pruning, and traversal, forming the algorithmic foundation for both parallel SGRAND and enhanced ORBGRAND.

Building on this unified framework, we develop a parallelized version of SGRAND, which concurrently generates and evaluates multiple EPs to achieve latency reduction while maintaining ML optimality. To guarantee ML decoding in the parallel setting, a newly designed early-termination criterion is proposed and theoretically proven to preserve optimality. Moreover, several tree-based acceleration techniques, such as pruning strategy, recursive reliability computation, and early termination rule, are incorporated to further reduce both the number of decoding tests and the computational complexity.

We further demonstrate that the EPs in ORBGRAND can also be embedded into the EP tree representation. Building on this result, we propose a hybrid enhanced ORBGRAND scheme: ORBGRAND is first executed to quickly identify a codeword, after which the proposed parallel SGRAND is invoked to determine the result as the ML solution. This hybrid design achieves ML decoding with only a small number of additional EP evaluations beyond standard ORBGRAND, while fully preserving ORBGRAND’s inherent parallel efficiency and achieving lower decoding latency compared to parallel SGRAND.

The main contributions of this work are summarized as follows:

(1) \textbf{Unified EP tree framework:} Proposes a unified representation that describes the structural relationship of EPs in both SGRAND and ORBGRAND, serving as a common basis for parallel exploration and algorithmic optimization.
    
(2) \textbf{Parallelized SGRAND:} Develops a parallel decoding method that generates and evaluates multiple EPs concurrently to reduce latency, supported by a new termination criterion that ensures ML optimality and tree-based techniques that lower decoding complexity.
    
(3) \textbf{Hybrid Enhanced ORBGRAND:} We show that the EPs of ORBGRAND can be embedded into the EP tree, and leverage parallel SGRAND as a post-processing mechanism to recover ML optimality with a controlled additional complexity.

We validate the proposed algorithms through theoretical analysis and numerical experiments. All simulations are evaluated using an operation-count--based effective complexity metric under an idealized parallel execution model. Taking serial SGRAND as the baseline, parallel SGRAND and the proposed ML-enhanced ORBGRAND achieve effective complexity reduction factors of $3.96\times$ and $4.21\times$, respectively.

The remaining part of this paper is organized as follows: Section \ref{sec:related} provides a brief survey of GRAND family algorithms, including their parallelization characteristics, performance trade-offs, and hardware realizations. Section \ref{sec:preliminaries} introduces some preliminaries, including the transmission model and the GRAND algorithm. Section \ref{sec:para_SGRAND} presents the EP tree, the tree-based representation of SGRAND and its parallelized design. Section \ref{sec:extend_SGRAND} further presents several acceleration techniques for the parallelized SGRAND. Section \ref{sec:orbgrand} turns to the development of the hybrid enhanced ORBGRAND. Section \ref{sec:simulation} exhibits results of numerical experiments. Section \ref{sec:conclusion} concludes the paper.

\section{Related Work}\label{sec:related}

The applicability of GRAND in next-generation communication systems has been actively investigated \cite{yue2023efficient,gautam2025analysis}. For example, \cite{yue2023efficient} evaluated decoding short codes using GRAND and Ordered Statistics Decoding (OSD) \cite{fossorier2002soft,yue2021probability}, showing its advantages at high code rates under favorable channel conditions. Similarly, \cite{gautam2025analysis} compared GRAND with several general-purpose decoders, including Automorphism Ensemble Decoding (AED) \cite{pillet2022classification}, OSD, Belief Propagation Decoding (BPD) \cite{luby2002improved}, and Bit-Flipping Decoding (BFD) \cite{afisiadis2014low}. The recent work \cite{wang2025guessing} further provides a unified perspective on guessing-based decoding, comparing GRAND and Guessing Codeword Decoding (GCD) under a common framework and highlighting their structural similarities and performance trade-offs.

Focusing on specific GRAND variants, the original hard-decision algorithms support Hamming weight ordered EP generation, allowing parallel exploration without losing ML optimality \cite{duffy2019capacity}, thereby enabling efficient hardware implementation \cite{abbas2020high}. For channels with soft output, SGRAND \cite{solomon2020soft} achieves ML decoding by dynamically maintaining a candidate set to generate EPs in a strictly descending order of posterior probabilities; however, existing methods rely on sequential generation and scheduling of EPs, which limits parallel execution and leads to increased decoding latency.

To address the complexity of EP generation and enable parallelism, ORBGRAND \cite{duffy2022ordered} and its ORB-type variants \cite{liu2022orbgrand,wan2024approaching} adopt a predefined ordering of EPs based solely on the ranking of the received channel reliability values. By eliminating dependencies among consecutive EPs, the resulting ORB-type GRAND algorithms are amenable to efficient parallel implementation, which has been extensively investigated in hardware \cite{abbas2023guessing, ji2024efficient, xiao2023low, 9715725, abbas2022high, abbas2025improved}.

Beyond these developments, several recent works have focused on improving decoding performance or extending the applicability of guessing-based decoding. In particular, multi-branch or list-based approaches generate multiple candidate EPs by enumerating configurations over a subset of unreliable positions, enabling parallel exploration of the search space \cite{zheng2024locally, zheng2024universal}. Hybrid decoding architectures have also been proposed, where a fast reliability-based search is followed by a refinement stage, such as algebraic decoding \cite{wang2025two}. While these approaches may allow certain degrees of parallelism, they generally do not guarantee strict ML optimality under parallel execution.

The drawback of ORB-type GRAND algorithms lies in their reliance on the ranking, rather than exact values, of channel reliability values. This drawback leads to suboptimal decoding performance. Although information-theoretic studies \cite{liu2022orbgrand,li2024orbgrand} show that ORBGRAND is almost capacity-achieving, it suffers a performance gap relative to SGRAND at finite codeword lengths, particularly in the low-block error rate (BLER) regime. Several refinements of the prescribed EP sets \cite{duffy2022ordered,liu2022orbgrand,condo2021iLWO,wan2024approaching} help reduce the gap but cannot eliminate it. Some alternative approaches attempt to bridge it by incorporating a small amount of channel reliability values \cite{wan2025finetuning} or outputting multiple candidates via listing decoding \cite{abbas2022list}; these introduce additional complexity and still cannot achieve ML decoding.

Taken together, a substantial body of work attempts to approach the EP ordering and decoding performance of SGRAND through efficient parallel implementations. Yet, the prevailing dilemma remains: SGRAND is ML-optimal but sequential, while ORBGRAND is parallelizable but suboptimal. This motivates our work in this paper to parallelize SGRAND preserving ML performance and to enhance ORBGRAND for achieving ML performance with minimal overhead.

\section{Preliminaries}\label{sec:preliminaries}

We use capital letters (e.g., $X$) to represent random variables and their corresponding lowercase letters (e.g., $x$) to represent the realizations of random variables.

\subsection{Transmission Model}\label{subsec:model}

We consider a general block coding scheme where information bits $\underline{U} \in \mathbb{F}_{2}^{K}$ are encoded into a codeword $\underline{W} \in \mathcal{C} \subseteq \mathbb{F}_{2}^{N}$ at code rate $R = \frac{K}{N}$. For analytical simplicity, we assume antipodal signaling over an additive white Gaussian noise (AWGN) channel, but the principles in our work can be readily applied to more general channels. For a given codeword, the transmitted vector $\underline{X}$ satisfies $X_i = 1$ if $W_i = 0$ and $X_i = -1$ if $W_i = 1$ for $i = 1, \ldots, N$. The corresponding channel output vector follows the conditional Gaussian distribution $\underline{Y}|\underline{X} \sim \mathcal{N}(\underline{X}, \sigma^2 \mathbf{I}_{N \times N})$.

Given the received signal vector $\underline{Y}$, the log-likelihood ratios (LLRs) are computed as:
\begin{equation*}\label{eq:LLR}
    \text{LLR}_{i} = \log\frac{p_{Y|W}(Y_i \mid W_{i} = 0)}{p_{Y|W}(Y_i \mid W_{i} = 1)} = \frac{2}{\sigma^2} Y_i, \quad i = 1, \ldots, N.
\end{equation*}
We denote $L_i = |\text{LLR}_i|$ and call it the reliability of the $i$-th channel output, with $\ell_i$ denoting the realization of $L_i$.

We define the hard decision function $\theta(\cdot)$ such that $\theta(y) = 0$ if $y \geq 0$ and $\theta(y) = 1$ otherwise. For the received vector $\underline{Y}$, we denote $\theta(\underline{Y}) = [\theta(Y_1), \ldots, \theta(Y_N)] \in \mathbb{F}_2^N$ for notational convenience. 

Following elementary calculation (see, e.g., \cite{wan2024approaching}), the conditional probability that a transmitted bit is incorrectly decided by the hard decision, conditioned upon a channel output, is given by $p(\theta(Y_i) \neq W_i \mid Y_i = y_i) = 1/[1 + \exp(\ell_i)]$,
from which we have a useful relationship for subsequent analysis, for $e \in \mathbb{F}_2$:
\begin{equation}\label{eq:p_ell}
    \frac{p_{W|Y}(\theta(y_i)\mid y_i)}{p_{W|Y}(e \oplus \theta(y_i)\mid y_i)} = \begin{cases}
        \exp(\ell_i), & \text{if } e = 1 \\
        1, & \text{if } e = 0
    \end{cases}.
\end{equation}

\subsection{GRAND}\label{subsec:GRAND}

GRAND is a recently proposed decoding paradigm that consists of two fundamental components: an EP generator and a codeword tester. In general, a GRAND algorithm operates through the following systematic workflow:

(1) Given a maximum query limit $T$, generate a sequence of EPs $\{\underline{e}(1),\dots,\underline{e}(T)\}$ where each EP is an element in $\mathbb{F}_{2}^{N}$.

(2) Sequentially test the EPs $\{\underline{e}(1), \ldots, \underline{e}(T)\}$. For $\underline{e}(t)$, verify whether $\theta(\underline{y}) \oplus \underline{e}(t)$ constitutes a valid codeword. Declare the first such identified codeword as the decoding result and terminate the workflow; or declare decoding failure if there is no valid codeword identified for all EPs.

This procedure is formalized in Algorithm \ref{alg:GRAND}. Different GRAND algorithms differ primarily in their EP generation strategies. The specific sequence of generated EPs directly affects the decoding performance. The following proposition establishes an EP generation strategy that achieves ML decoding.

\begin{algorithm}[tbp]
    \caption{General workflow of GRAND}\label{alg:GRAND}
    \textbf{Inputs}: $\underline{y}$, $T$\;
    \textbf{Outputs}: $\hat{\underline{w}}$\;
    Initialize $t = 0$ and $Tag = 0$\;
    \While{$Tag = 0$ \textbf{AND} $t < T$}{
        $t = t + 1$\;
        $\underline{e}(t) \gets $ EP\_Generator($\underline{y}, t$)\;
        Tag $\gets$ Codeword\_Tester($\theta(\underline{y})\oplus \underline{e}(t)$)\tcp*{Tag = 1 if successful, otherwise 0}
    }
    \eIf{$Tag = 1$}{
        $\hat{\underline{w}} = \theta(\underline{y})\oplus \underline{e}(t)$\;
    }{
        $\hat{\underline{w}} = \emptyset$ \tcp*{Decoding failure}
    }
\end{algorithm}

\begin{proposition}\label{prop:equal_ml}
    We define the soft weight of an EP $\underline{e}$ as $\zeta(\underline{e}) = \sum_{i:e_i = 1} \ell_i$. If we set the maximum query limit to $T = 2^N$ and generate $\underline{e}(t)$ for $1 \leq t \leq 2^N$ such that the sequence $\{\zeta(\underline{e}(t))\}_{t = 1,\ldots,2^N}$ is monotonically non-decreasing, then the resulting GRAND algorithm produces an ML-optimal codeword.
\end{proposition}

\begin{IEEEproof}
This result has been well established in the literature; see, e.g., \cite{dorsch1974decoding, solomon2020soft}. For completeness, we provide a brief derivation highlighting the key steps.

\vspace{-0.3cm}
{\small
\begin{align}
    \hat{\underline{w}} 
    & = \arg\max_{\underline{w}\in \mathcal{C}}\left\{p_{\underline{Y}|\underline{W}}(\underline{y} \mid \underline{w})\right\} = \arg\min_{\underline{w}\in \mathcal{C}}\left\{\frac{p_{\underline{W}|\underline{Y}}(\theta(\underline{y}) \mid \underline{y})}{p_{\underline{W}|\underline{Y}}(\underline{w} \mid \underline{y})}\right\} \label{eq:ml_0} \\
    & = \theta(\underline{y}) \oplus \arg\min_{\underline{e}:\underline{e} \oplus \theta(\underline{y})\in \mathcal{C}}\left\{\frac{p_{\underline{W}|\underline{Y}}(\theta(\underline{y}) \mid \underline{y})}{p_{\underline{W}|\underline{Y}}(\underline{e}  \oplus \theta(\underline{y}) \mid \underline{y})}\right\}  \label{eq:ml_1} \\
    & = \theta(\underline{y}) \oplus \arg\min_{\underline{e}:\underline{e} \oplus \theta(\underline{y})\in \mathcal{C}}\left\{\sum_{i:e_i = 1} \ell_i \right\} \label{eq:ml_2} \\
    & = \theta(\underline{y}) \oplus \arg\min_{\underline{e}:\theta(\underline{y})\oplus \underline{e}\in \mathcal{C}} \left\{\zeta(\underline{e})\right\}. \label{eq:ml_3}
\end{align}
}

The derivation follows by substituting $\underline{w} = \theta(\underline{y}) \oplus \underline{e}$ into \eqref{eq:ml_0} to get \eqref{eq:ml_1}, and applying \eqref{eq:p_ell} to obtain \eqref{eq:ml_2}.
\end{IEEEproof}

The final formulation in \eqref{eq:ml_3} establishes that the optimal EP minimizes the soft weight $\zeta(\underline{e})$ among all EPs that produce valid codewords in $\mathcal{C}$. This is exactly what SGRAND does, generating EPs in ascending order of their soft weights, and therefore SGRAND identifies the optimal EP that solves \eqref{eq:ml_3}, thereby achieving ML decoding.

For the remainder of this paper, we introduce the following terminologies:
\begin{itemize}
    \item An EP $\underline{e}$ is \emph{valid} if it satisfies $\theta(\underline{y})\oplus \underline{e} \in \mathcal{C}$.
    \item A valid EP $\underline{e}$ is \emph{optimal} if it satisfies\\ $\underline{e} = \arg\min_{\underline{e}:\theta(\underline{y})\oplus \underline{e}\in \mathcal{C}} \left\{\zeta(\underline{e})\right\}$.
    \item During the execution of GRAND, a valid EP $\underline{e}(t')$ is \emph{locally optimal} if it satisfies:
    \begin{equation}\label{eq:local_opti}
        \zeta(\underline{e}(t')) = \min_{\underline{e}(t):\theta(\underline{y})\oplus \underline{e}(t) \in \mathcal{C}, 1\leq t\leq t'}\{\zeta(\underline{e}(t)) \}.
    \end{equation}
\end{itemize}


\section{EP Tree Representation and Parallelization of SGRAND}\label{sec:para_SGRAND}

According to Proposition \ref{prop:equal_ml}, SGRAND proposed in \cite{solomon2020soft} implements ML decoding.\footnote{Strictly speaking, this assertion holds only when no abandonment is applied, i.e., $T = 2^N$.} In order to parallelize SGRAND, we utilize a binary tree structure to represent EPs and describe SGRAND based on this structure, leading to a parallelized SGRAND implementation.

\subsection{Error Pattern Tree}

To enable systematic parallel exploration of EPs, we introduce a binary tree structure that organizes all possible EPs according to their reliability ranking relationship. This tree explicitly formalizes the implicit search structure induced by SGRAND \cite{solomon2020soft}. Similar tree-based ideas have appeared in other decoding frameworks, such as OSD \cite{yue2021probability, liang2024random, zheng2024universal}, where carefully designed partial-order relationships are used to organize EPs and guide the search process. In contrast, the proposed EP tree provides an explicit structural representation of the search process in SGRAND, making the ordering induced by reliability more amenable to systematic exploration and parallelization.

\begin{definition}[EP tree]\label{def:tree}
An EP tree is a binary tree in which each node corresponds to a unique EP, i.e., a binary vector in $\mathbb{F}_2^N$.
For a received $\underline{\ell}$, let $\underline{r}=(r_1,\ldots,r_N)$ be a permutation of $\{1,\ldots,N\}$ that sorts the entries of $\underline{\ell}$ in nondecreasing order: $\ell_{r_1} \leq \ell_{r_2} \le \cdots \le \ell_{r_N}$. Based on $\underline{r}$, the EP tree satisfies the following structural properties:
\begin{itemize}
    \item The root node represents the all-zero vector $\underline{0}$.
    \item The root node has a single child corresponding to the pattern with $e_{r_1} = 1$ and all other components being $0$.
    \item For any non-zero node $\underline{e}$, define $j^{*} = \max_j\{e_{r_j} = 1\}$. If $j^{*} = N$, the node is a leaf with no children; otherwise, $j^{*} \neq N$, the node has two children, denoted as left child $\underline{e}_{L}$ and right child $\underline{e}_{R}$:
    \begin{enumerate}
        \item $\underline{e}_{L}\leftarrow \underline{e},\text{ and set }\ e_{L,r_{j^{*}}} = 0,\ e_{L,r_{j^{*}+1}} = 1$. \label{eq:left_child}
        \item $\underline{e}_{R}\leftarrow \underline{e},\text{ and set }\ e_{R,r_{j^{*}+1}} = 1$.\label{eq:right_child}
    \end{enumerate}
\end{itemize}
\end{definition}

The depth of a node refers to the length of the path from the root node to that node. Consequently, the root node has depth $0$, and the depth of any node is $\max_j\{e_{r_j} = 1\}$.

\subsection{Revisiting SGRAND}\label{subsec:revisit_SGRAND}

Having established the EP tree, we can reinterpret SGRAND \cite{solomon2020soft} as a best-first traversal: starting from the root node, the decoder maintains a frontier $\mathcal{S}$ of candidate nodes; at each iteration it removes the node with the minimum $\zeta(\cdot)$, tests the corresponding EP for codeword validity, and inserts its children into $\mathcal{S}$. This best-first expansion visits EPs in nondecreasing $\zeta$-order and therefore achieves ML optimality.

This tree-based interpretation can be formalized as a systematic tree traversal process:

(1) Initialize $\mathcal{S} = \{\underline{0}\}$ as the candidate set, set iteration counter $t = 0$, sort $\underline{\ell}$ to obtain $\underline{r}$.

(2) Increment $t$ by 1. Remove from $\mathcal{S}$ the EP $\underline{e}$ with the smallest $\zeta(\underline{e})$, and add its children to $\mathcal{S}$.

(3) If $\theta(\underline{y}) \oplus \underline{e}$ is a valid codeword, output it as the decoding result; otherwise, if $t \leq T$, return to step 2, and if $t > T$, declare decoding failure.

This process is essentially the algorithm in \cite{solomon2020soft} restated in terms of EP tree, as described in Algorithm \ref{alg:SGRAND_tree}, which can be used as the EP Generator module in Algorithm \ref{alg:GRAND}. We illustrate the algorithm through a concrete example.

\begin{algorithm}[htbp]\label{alg:SGRAND_tree}
    \caption{EP Generator for SGRAND \cite{solomon2020soft}}
    \textbf{Inputs}: $\underline{y}$, $t$, $\mathcal{S}$, $\underline{r}$\;
    \textbf{Outputs}: $\underline{e}(t)$ \tcp*{Line 6 in Algorithm \ref{alg:GRAND}}
    \If{$t = 1$}{
        $\mathcal{S} \gets \{\underline{0}\},\ \underline{r}\gets $ reliability ordering vector
    }
    $\underline{e}(t) \gets \arg\min_{\underline{e}\in \mathcal{S}}\zeta(\underline{e})$ \tcp*{The $t$-th EP}
    $\mathcal{S} = \mathcal{S} \backslash \{\underline{e}(t)\}$ \tcp*{Lines 7-14: Update $\mathcal{S}$}
    $j^{*} = 
    \begin{cases} 
        0 & \text{if } \underline{e}(t) = \underline{0}\\
        \max_{j}\{e_{r_j}(t) = 1\} & \text{otherwise}
    \end{cases}$\;
    \If{$j^{*} < N$}{
        $e_{r_{j^{*}+1}} \gets 1 $, $\mathcal{S} = \mathcal{S}\cup \{\underline{e}\}$\;
        \If{$j_{*} > 0$}{
            $e_{r_{j^{*}}} \gets 0$, $ \mathcal{S} = \mathcal{S}\cup \{\underline{e}\}$\;
        }
    }
\end{algorithm}

\begin{example}\label{ex:tree_sgrand}
    Consider a received vector of length $N=4$ with reliability weights $\underline{\ell} = (0.8, 1.2, 2.1, 3.4)$, which yields $\underline{r}=(1,2,3,4)$. Following Definition \ref{def:tree}, we construct the corresponding EP tree as illustrated in Figure \ref{fig:ex_tree}.
    
    To visualize the algorithm's execution, we go through the decoding process for $t=5$ iterations. Table \ref{tab:error_patterns} records the evolution of $\mathcal{S}$. The state at $t=5$ is graphically represented in Figure \ref{fig:ex_tree}, where red nodes highlight the currently maintained EPs in $\mathcal{S}$, black nodes are EPs that have been evaluated, and gray nodes represent unexplored EPs.

\begin{table}[htbp]
    \centering
    \renewcommand{\arraystretch}{1.4}
    \begin{tabular}{|c|c|c|c|}
    \hline
    $t$ & $\underline{e}(t)$ & $\zeta(\underline{e}(t))$ & $\mathcal{S}$ \\ \hline
    $1$ & $(0000)$  & $0$ & $\{(1000)\}$  \\ \hline
    $2$ & $(1000)$  & $0.8$ &  $\{(0100), (1100)\}$ \\ \hline
    $3$ & $(0100)$  & $1.2$ &  $\{(0010), (0110), (1100)\}$ \\ \hline
    $4$ & $(1100)$  & $2.0$ &  $\{(0010), (0110), (1010), (1110)\}$ \\ \hline
    $5$ & $(0010)$  & $2.1$ &  $\{(0110), (1010), (1110), (0001), (0011)\}$ \\ \hline
    \end{tabular}
    \\[5pt]
    \caption{Example \ref{ex:tree_sgrand}: EP $\underline{e}(t)$, $\zeta$ value and EP set $\mathcal{S}$ at each iteration $t$.}
    \label{tab:error_patterns}
\end{table}

\begin{figure}[htbp]
    \centering
    \resizebox{0.48\textwidth}{!}{%
    \begin{tikzpicture}
    \tikzset{level distance=19pt, sibling distance=1pt, dashed edge/.style={edge from parent/.style={dashed,draw}}}
    \Tree
    [.0000
        [.1000
            [.0100
                    [.0010
                        [.\textcolor{red}{0001} ]
                        [.\textcolor{red}{0011} ]
                    ]
                    [.\textcolor{red}{0110} 
                        \edge[dashed];[.\textcolor{gray}{0101} ]
                        \edge[dashed];[.\textcolor{gray}{0111} ]
                    ]
                ]
            [.1100
                [.\textcolor{red}{1010} 
                    \edge[dashed];[.\textcolor{gray}{1001} ] 
                    \edge[dashed];[.\textcolor{gray}{1011} ]
                ]
                [.\textcolor{red}{1110}  
                    \edge[dashed];[.\textcolor{gray}{1101} ]
                    \edge[dashed];[.\textcolor{gray}{1111} ]
                ]
            ]
        ]
    ]
    \end{tikzpicture}
    }
    \caption{Example \ref{ex:tree_sgrand}: EP tree when $\underline{r}=(1,2,3,4)$. Red nodes: $\mathcal{S}$ at $t=5$; black nodes: examined; gray nodes: not yet retrieved.}
    \label{fig:ex_tree}
\end{figure}
\end{example}

\subsubsection{Implementation and Complexity} The computational complexity of Algorithm \ref{alg:GRAND} is dominated by two primary operations: EP generation and codeword verification. For SGRAND, the complexity of EP generation stems from:
\begin{enumerate}
    \item Selecting the optimal $\underline{e}$ and removing it from $\mathcal{S}$.
    \item Constructing new EPs and inserting them into $\mathcal{S}$.
\end{enumerate}
These operations must be executed at each iteration $t$. If an array is used to store $\mathcal{S}$, the complexity of selecting $\underline{e}$ is $\mathcal{O}(t)$, leading to an overall complexity of $\mathcal{O}(t^2)$. We denote the complexity at iteration $t$ as $\mathcal{T}(t)$. For the vanilla implementation using arrays, the complexity is:

\vspace{-0.3cm}
{\small
\begin{equation}\label{eq:complexity_array_sum}
\mathcal{T}_{\text{array}}(t)
=  \underbrace{\mathcal{O}(t^2)+\mathcal{O}(t)}_{\text{select \& remove}}+\underbrace{\mathcal{O}(Nt) + \mathcal{O}(t)}_{\text{construct \& insert}} + \underbrace{t \cdot f(N,K)}_{\text{test}},
\end{equation}
}

\noindent where $f(N,K)$ is the codeword testing complexity. In \cite{solomon2020soft}, it is proposed to use a min-heap data structure to implement $\mathcal{S}$.\footnote{A min-heap is a complete binary tree where each node's value is no greater than those of its children \cite[Chapter 6]{cormen2022introduction}.} This approach reduces the complexity of EP selection to $\mathcal{O}(1)$, while maintaining the heap after deletion requires $\mathcal{O}(\log t)$. The complexity of generating and inserting subsequent EPs are $\mathcal{O}(N)$ and $\mathcal{O}(\log t)$, respectively. By Stirling's formula, we have $\sum_{i=1}^{t}\log i \sim \mathcal{O}(t\log t)$, leading to:

\vspace{-0.3cm}
{\small
\begin{equation}\label{eq:complexity_heap_sum}
\mathcal{T}_{\text{heap}}(t)
=  \underbrace{\mathcal{O}(t)+\mathcal{O}(t\log t)}_{\text{select \& remove}}+\underbrace{\mathcal{O}(Nt) + \mathcal{O}(t\log t)}_{\text{construct \& insert}} + \underbrace{t \cdot f(N,K)}_{\text{test}}.
\end{equation}
}

The expressions in \eqref{eq:complexity_array_sum} and \eqref{eq:complexity_heap_sum} characterize the asymptotic growth behavior of SGRAND implementations with respect to the query number $t$. In contrast, the complexity metric $\Phi$ introduced in Section \ref{sec:simulation} provides an instantiated, operation-count-based evaluation under specific system parameters.

\begin{example}\label{ex:complexity_heap_sum}
    Continuing Example \ref{ex:tree_sgrand}, assume decoding has not terminated at $t = 5$ and the algorithm proceeds to iteration $t = 6$. Figure \ref{fig:tree-1} displays the min-heap structure representing the current candidate set $\mathcal{S}$. The subsequent operations are also given in Figure \ref{fig:heap-example}.
\end{example}

\begin{remark}
    In Examples \ref{ex:tree_sgrand} and \ref{ex:complexity_heap_sum}, there are two tree structures. Figure \ref{fig:ex_tree} shows the EP tree, which includes all EPs in $\mathbb{F}^N_2$. At each iteration $t$, $\mathcal{S}$ corresponds to a part of nodes. In order to efficiently store and update $\mathcal{S}$ in the implementation of algorithm, the min-heap structure shown in Figure \ref{fig:heap-example} is adopted.
\end{remark}


\begin{figure*}[htbp]
    \centering
    \subfigure[Initial heap at $t=6$.]{
      \resizebox{!}{0.13\textheight}{
        \begin{forest}
          for tree={
            circle, draw, minimum size=2.5em,
            inner sep=1pt, font=\scriptsize,
            s sep=7mm, edge={-, semithick},
            align=center,
          },
          [{\shortstack{2.9 \\ \texttt{1010}}}
            [{\shortstack{3.3 \\ \texttt{0110}}}
              [{\shortstack{3.4 \\ \texttt{0001}}}]
              [{\shortstack{4.1 \\ \texttt{1110}}}]
            ]
            [{\shortstack{5.5 \\ \texttt{0011}}}]
          ]
        \end{forest}}
      \label{fig:tree-1}
    }
    \subfigure[Delete the root node and perform maintenance.]{
      \resizebox{!}{0.13\textheight}{
        \begin{forest}
            for tree={
              circle, draw, minimum size=2.5em,
              inner sep=1pt, font=\scriptsize,
              s sep=7mm, edge={-, semithick},
              align=center,
            },
            [{Delete}, name=root
              [{\shortstack{3.3 \\ \texttt{0110}}}, name=left
                [{\shortstack{3.4 \\ \texttt{0001}}}, name=leftleft]
                [{\shortstack{4.1 \\ \texttt{1110}}}, name=leftright]
              ]
              [{\shortstack{5.5 \\ \texttt{0011}}}, name=right]
            ]
            \draw[->, thick] (left) to[out=135,in=180] node[midway, above, sloped] {\scriptsize Move up} (root);
            \draw[->, thick] (leftleft) to[out=135,in=180] node[midway, above, sloped] {\scriptsize Move up} (left);
          \end{forest}}
      \label{fig:tree-2}
    }
    \subfigure[The structure after deletion.]{
      \resizebox{!}{0.13\textheight}{
        \begin{forest}
          for tree={
            circle, draw, minimum size=2.5em,
            inner sep=1pt, font=\scriptsize,
            s sep=7mm, edge={-, semithick},
            align=center,
          },
          [{\shortstack{3.3 \\ \texttt{0110}}}
            [{\shortstack{3.4 \\ \texttt{0001}}}
              [{\shortstack{4.1 \\ \texttt{1110}}}]
            ]
            [{\shortstack{5.5 \\ \texttt{0011}}}]
          ]
        \end{forest}}
      \label{fig:tree-3}
    }
    \subfigure[Insert the first child node and performance maintenance.]{
      \resizebox{!}{0.13\textheight}{
        \begin{forest}
          for tree={
            circle, draw, minimum size=2.5em,
            inner sep=1pt, font=\scriptsize,
            s sep=7mm, edge={-, semithick},
            align=center,
          },
          [{\shortstack{3.3 \\ \texttt{0110}}}, name=root
            [{\shortstack{3.4 \\ \texttt{0001}}}, name=left
              [{\shortstack{4.1 \\ \texttt{1110}}}, name=leftleft]
              [{\shortstack{4.2 \\ \texttt{1001}}}, name=leftright]
            ]
            [{\shortstack{5.5 \\ \texttt{0011}}}]
          ]
          \draw[-, thick] (left) to[out=0,in=45] node[pos=0.7, right=2pt] {\scriptsize Compare} (leftright);
        \end{forest}}
      \label{fig:tree-4}
    }
    \subfigure[The heap after $t=6$.]{
      \resizebox{!}{0.13\textheight}{
        \begin{forest}
          for tree={
            circle, draw, minimum size=2.5em,
            inner sep=1pt, font=\scriptsize,
            s sep=7mm, edge={-, semithick},
            align=center,
          },
          [{\shortstack{3.3 \\ \texttt{0110}}}
            [{\shortstack{3.4 \\ \texttt{0001}}}
              [{\shortstack{4.1 \\ \texttt{1110}}}]
              [{\shortstack{4.2 \\ \texttt{1001}}}]
            ]
            [{\shortstack{5.5 \\ \texttt{0011}}}
              [{\shortstack{6.3 \\ \texttt{1011}}}]
            ]
          ]
        \end{forest}}
      \label{fig:tree-5}
    }
    \caption{Operations on the min-heap at $t=6$ (continuing Example \ref{ex:tree_sgrand}).}
    \label{fig:heap-example}
  \end{figure*}


\subsection{Parallel SGRAND}\label{subsec:basic-parallel-SGRAND}

The basic approach to parallelizing SGRAND involves selecting multiple EPs from the candidate set $\mathcal{S}$ for concurrent testing, followed by the insertion of their respective child nodes into $\mathcal{S}$. The tree-based representation of EPs provides a key structural foundation that guarantees complete and non-redundant traversal of the search space when this process is executed iteratively.

Our proposed parallel SGRAND algorithm operates according to the following methodology, illustrated schematically in Figure \ref{fig:basic_method}:
\begin{itemize}
    \item Step 1: Compute $\underline{r}$. Initialize the round counter $k = 0$ and the candidate set $\mathcal{S}_{0} = \{\underline{0}\}$.
    \item Step 2: Increment $k \leftarrow k + 1$. Extract $P_{k}$ EPs with the smallest $\zeta(\cdot)$ from $\mathcal{S}_{k-1}$ for concurrent checking. Here for each $k$, $P_{k} \leq |\mathcal{S}_{k-1}|$ is a parameter controlling the degree of parallelism. Denote this test batch as:
    \begin{equation}
        \mathcal{E}_{k} = \left\{\underline{e}\left(\sum_{i=1}^{k-1}P_{i} + 1\right),...,\underline{e}\left(\sum_{i=1}^{k}P_{i}\right)\right\}.
    \end{equation} 
         Generate the child nodes of $\mathcal{E}_{k}$ and update the candidate set to $\mathcal{S}_{k} = (\mathcal{S}_{k-1}\backslash \mathcal{E}_{k}) \cup \mathcal{E}_{k}^+$, where $\mathcal{E}_{k}^+$ denotes the child nodes of EPs in $\mathcal{E}_{k}$.
    \item Step 3: Test all EPs in $\mathcal{E}_{k}$ concurrently. If no valid EPs are found, return to Step 2; if some valid EPs exist, identify the EP $\underline{e}^{*}$ with the minimum $\zeta(\cdot)$ and proceed to Step 4.
\end{itemize}

\begin{figure}[tbp]
\centering
\begin{tikzpicture}[
    scale=0.95,
    set/.style={
        circle,
        draw,
        minimum size=1.2cm,
        align=center,
        font=\small,
        fill=white
    },
    checker/.style={
        rectangle,
        draw,
        minimum width=2.5cm,
        minimum height=0.4cm,
        align=center,
        fill=blue!8,
        inner sep=1pt
    },
    checkerGroup/.style={
        rectangle,
        draw,
        rounded corners=5pt,  
        inner xsep=4pt,     
        inner ysep=4pt,       
        fill=blue!3,
        dashed
    },
    arrow/.style={
        ->, 
        >=stealth, 
        thick,
        shorten >=2pt,
        shorten <=2pt
    },
    time/.style={
        font=\footnotesize,
        above right=-0.3cm and 0.1cm,
        text=gray
    },
    scale=0.88,
    transform shape
]

\node[set, fill=yellow!20] (s0) at (-2.5, 0) {$\mathcal{S}_0$};
\draw[dashed, gray] (-1.85, -2.5) -- (-1.85, 1) node[time] {${k=1}$};

\node[set] (e1) at (0.25, 0) {$\mathcal{S}_0 \backslash\mathcal{E}_{1}$};
\node[set, fill=green!15] (e1d) at (-1.0, -1.7) {$\mathcal{E}_{1}$};

\foreach \i in {0,...,3} {
    \node[checker] (checker1-\i) at (0.1, -3.4 - \i*0.5) {Tester \pgfmathparse{int(\i+1)}\pgfmathresult};
}
\node[text=blue!50, font=\tiny] at (0, -5.5) {Parallel Codeword Testers (4 units)};

\begin{scope}[on background layer]
    \node[checkerGroup, fit=(checker1-0)(checker1-3)(checker1-3.south east)] {};
\end{scope}

\node[set] (e1child) at (1.2, -1.7) {$\mathcal{E}_{1}^{+}$};
\node[set, fill=yellow!20] (s1) at (3, 0) {$\mathcal{S}_1$};

\draw[arrow] (s0) -- (e1);
\draw[arrow] (s0) -- node[right, pos=0.4] {top $P_1$} (e1d);
\draw[arrow] (e1d) -- (-1.0, -3.0);
\draw[arrow] (1.2, -3.0) -- (e1child);
\draw[arrow, thick] (e1child) -- (s1);
\draw[arrow] (e1) -- (s1);

\draw[dashed, gray] (3.65, -2.5) -- (3.65, 1) node[time] {${k=2}$};

\node[set] (e2) at (5.75, 0) {$\mathcal{S}_1 \backslash\mathcal{E}_{2}$};
\node[set, fill=green!15] (e2d) at (4.5, -1.7) {$\mathcal{E}_{2}$};

\foreach \i in {0,...,3} {
    \node[checker] (checker2-\i) at (5.5, -3.4 - \i*0.5) {Tester \pgfmathparse{int(\i+1)}\pgfmathresult};
}
\node[text=blue!50, font=\tiny] at (5.5, -5.5) {Parallel Codeword Testers (4 units)};

\begin{scope}[on background layer]
    \node[checkerGroup, fit=(checker2-0)(checker2-3)(checker2-3.south east)] {};
\end{scope}

\node at (5.75, -1.7) {$\cdots$};
\node at (6.75, 0) {$\cdots$};

\draw[arrow] (s1) -- (e2);
\draw[arrow] (s1) -- node[right, pos=0.4] {top $P_2$}(e2d);
\draw[arrow] (e2d) -- (4.5, -3.0);

\end{tikzpicture}
\caption{Schematic diagram of parallel SGRAND algorithm}
\label{fig:basic_method}
\end{figure}

A critical observation is that the EP $\underline{e}^{*}$ identified in Step 3 may represent only a locally optimal solution (as defined in \eqref{eq:local_opti}) rather than the optimal EP achieving ML decoding. This phenomenon arises because, for any two candidate sets $\mathcal{E}_{k_1}$ and $\mathcal{E}_{k_2}$ where $k_1 < k_2$, the condition
\begin{equation}\label{eq:k1_and_k2}
    \zeta(\underline{e}(i)) \leq \zeta(\underline{e}(j)), \ \forall\ \underline{e}(i) \in \mathcal{E}_{k_1},\ \underline{e}(j) \in \mathcal{E}_{k_2}
\end{equation}
cannot be guaranteed in general, except for the special case where $P_k = 1, \forall k$, i.e., the conventional serial implementation of SGRAND. To ensure that $\underline{e}^{*}$ achieves the optimal decoding, we need to continue with the following steps:
\begin{itemize}
    \item Step 4: Increment $k \leftarrow k + 1$. Generate $\mathcal{E}_k$ and $\mathcal{S}_k$ as in Step 2. Search for $\underline{e} \in \mathcal{E}_{k}$ that satisfies both $\theta(\underline{y}) \oplus \underline{e} \in \mathcal{C}$ and $\zeta(\underline{e}) < \zeta(\underline{e}^{*})$. If at least one such EPs exist, update $\underline{e}^{*}$ to the one with the minimum $\zeta(\cdot)$. Proceed to Step 5.
    \item Step 5: Determine $\tau_{k} = \min_{\underline{e}:\underline{e} \in \mathcal{S}_{k}}\{\zeta(\underline{e})\}$. If 
    $\tau_{k} \leq  \zeta(\underline{e}^{*})$, return to Step 4; otherwise, terminate decoding and output $\theta(\underline
    {y}) \oplus \underline{e}^{*}$.
\end{itemize}

The validity of the preceding algorithm for implementing parallel SGRAND is established through the following theorem:

\begin{theorem}\label{th:1}
    When the EP tree is complete with all the $2^N$ possible EPs, the parallel SGRAND algorithm implementing Steps 1-5 achieves ML decoding.
\end{theorem}
\begin{IEEEproof}
    We prove the assertion by revealing the monotonicity property of $\tau_{k}$:
    \begin{align}
        \tau_{k} & = \min_{\underline{e}:\underline{e} \in \mathcal{S}_{k}}\{\zeta(\underline{e})\} \notag\\
        & = \min\left\{\min_{\underline{e}:\underline{e} \in \mathcal{E}_{k+1}}\{\zeta(\underline{e})\},\min_{\underline{e}:\underline{e} \in \mathcal{S}_{k}\backslash \mathcal{E}_{k+1} }\{\zeta(\underline{e})\}\right\} \label{eq:th-1.1}\\ 
        & \leq \min\left\{\min_{\underline{e}:\underline{e} \in \mathcal{E}_{k+1}^{+}}\{\zeta(\underline{e})\},\min_{\underline{e}:\underline{e} \in \mathcal{S}_{k}\backslash \mathcal{E}_{k+1} }\{\zeta(\underline{e})\}\right\} \label{eq:th-1.2}\\
        & = \min_{\underline{e}:\underline{e} \in \mathcal{S}_{k+1}}\{\zeta(\underline{e})\} = \tau_{k+1}.
    \end{align}
    Herein, \eqref{eq:th-1.1} and \eqref{eq:th-1.2} analyze the transition from $\mathcal{S}_{k}$ to $\mathcal{S}_{k+1}$ through the removal of $\mathcal{E}_{k+1}$ and insertion of its child nodes, $\mathcal{E}_{k+1}^{+}$. When $\mathcal{E}_{k+1} = \emptyset$, \eqref{eq:th-1.2} trivially holds; otherwise, \eqref{eq:th-1.2} holds due to:
    \begin{align}
        \min_{\underline{e}:\underline{e} \in \mathcal{E}_{k+1}}\zeta(\underline{e}) &  = \min_{\underline{e}:\underline{e} \in \mathcal{E}_{k+1}}\left\{\min_{\underline{e}:\text{ with child}}\zeta(\underline{e}), \min_{\underline{e}:\text{ without child}}\zeta(\underline{e})\right\} \notag \\
        & \leq \min_{\underline{e}:\underline{e} \in \mathcal{E}_{k+1}\text{, with child}}\zeta(\underline{e}) < \min_{\underline{e}:\underline{e} \in \mathcal{E}_{k+1}^{+}}\zeta(\underline{e}).
    \end{align}

    During the decoding process, the algorithm discovers a sequence of locally optimal EPs $\{\underline{e}(t_1),\underline{e}(t_2),\ldots\}$, for some $t_1 < t_2 < \cdots$, satisfying:
    \begin{equation}
        \zeta(\underline{e}(t_1)) > \zeta(\underline{e}(t_2)) > \cdots. 
    \end{equation}

    Let $k_{t}$ denote the termination round.\footnote{If the termination condition is never met during computation, all EPs have been examined, thereby ensuring ML optimality.} At termination, the currently best EP $\underline{e}^{*}$ satisfies $\tau_{k_{t}-1} \leq \zeta(\underline{e}^{*}) < \tau_{k_{t}}$. We now argue that $\underline{e}^{*}$ solves $\arg\min_{\underline{e}:\theta(\underline{y})\oplus \underline{e}\in \mathcal{C}} \{\zeta(\underline{e})\}$.
    
    Suppose that there exists a better solution $\hat{\underline{e}}$ discovered at round $k_{t} + k'$ for some $k' > 0$ with $\zeta(\hat{\underline{e}}) < \zeta(\underline{e}^{*})$. This leads to a contradiction due to the chain of inequalities:
    \begin{equation}
        \zeta(\hat{\underline{e}}) \geq \tau_{k_{t}+k'} \geq \tau_{k_{t}} > \zeta(\underline{e}^{*}).
    \end{equation}
    Since the EP tree provides a complete coverage of $\mathbb{F}_2^N$, this contradiction establishes the ML optimality of $\underline{e}^{*}$.
\end{IEEEproof}

Theorem \ref{th:1} establishes that the proposed parallel SGRAND algorithm successfully maintains ML optimality while enabling parallel EP testing. The algorithm operates through iterative rounds during which multiple EPs are selected and tested concurrently. At each round $k$, the algorithm extracts $P_k$ EPs with the minimum soft weights from the current candidate set $\mathcal{S}_{k-1}$ for parallel testing. When valid EPs are discovered, the algorithm records the locally optimal solutions and continues searching until the termination condition $\tau_k > \zeta(\underline{e}^*)$ is satisfied, where $\tau_k$ represents the minimum soft weight among remaining candidates.

\subsection{Analysis and Design Considerations}

The most notable difference between parallel and serial implementations of SGRAND is the relaxation of the strict ordering constraint: while serial SGRAND tests EPs in exactly ascending order of $\zeta(\cdot)$, the parallel SGRAND processes EPs in batches where only the minimum $\zeta(\cdot)$ of each batch is required to satisfy the monotonicity property $\tau_{k-1} \leq \tau_k$. This relaxation enables concurrent processing but necessitates additional verification mechanisms to ensure global optimality, as done by Steps 4 and 5 in the previous subsection.

The batch size parameter $P_k$ serves as the key design lever for balancing parallelism and computational complexity. Small batch sizes closely resemble the behavior of serial SGRAND, preserving the ascending order of $\zeta(\cdot)$ to some extent, but reap limited parallelization gains. In contrast, large batch sizes fully realize parallelism but may require evaluating additional EPs.

While the proposed parallel SGRAND successfully achieves ML decoding, its computational complexity can be reduced via further optimization. The following section introduces several acceleration techniques, including pruning, recursive tree-based computation of reliability and parity, and early termination. These substantially reduce both the average number of EP evaluations and the per-evaluation computational complexity, while still maintaining ML guarantees.

\section{Accelerating Parallel SGRAND}\label{sec:extend_SGRAND}

Building upon the basic parallel SGRAND algorithm proposed in the previous section, in this section we present several acceleration techniques to further improve the performance of parallel SGRAND, while still maintaining the ML decoding performance.

\subsection{Pruning Strategy to Eliminate Unnecessary Tests}\label{subsec:4-1}

Our first acceleration technique is derived from the key observation that once a valid EP $\underline{e}^{*}$ is identified, all EPs with larger $\zeta(\cdot)$ can be safely discarded without compromising the ML optimality. The resulting pruning strategy exhibits significant computational benefits since eliminating a node in the EP tree automatically removes all its descendants from the search space.

In the original algorithm described in the previous section, we update $\mathcal{S}_{k}$ as $\mathcal{S}_{k} \gets (\mathcal{S}_{k-1}\backslash \mathcal{E}_{k}) \cup (\mathcal{E}_{k}^{+})$. We can modify the updating rule as follows:
\begin{enumerate}
    \item If $\exists\ \underline{e}\in \mathcal{E}_k$ satisfies $ \theta(\underline{y})\oplus \underline{e} \in \mathcal{C} $, then $\mathcal{S}_{k} \gets \emptyset$, otherwise $\mathcal{S}_{k} \gets \mathcal{S}_{k-1}\backslash \mathcal{E}_{k}$.
    \item For each $\underline{e} \in \mathcal{E}_{k}^{+}$, if $\zeta(\underline{e})>\zeta(\underline{e}^{*})$, it will not be added to $\mathcal{S}_k$, otherwise, $\mathcal{S}_k \gets \mathcal{S}_k\cup \{\underline{e}\}$.
\end{enumerate}
This systematic pruning strategy operates on both components of $\mathcal{S}_k$: (1) the remaining EPs in $\mathcal{S}_{k-1}\backslash \mathcal{E}_{k}$, and (2) the newly generated candidate EPs $\mathcal{E}_{k}^{+}$. Clearly, this strategy achieves substantial computational savings while maintaining the ML decoding performance.

\subsection{Leveraging Tree Structure to Accelerate Computation}\label{subsec:4-2}

Testing of each EP involves several computational tasks: selecting $\mathcal{E}_{k}$ from $\mathcal{S}_{k-1}$, verifying codeword validity, and calculating $\zeta(\cdot)$ for its child nodes. Accomplishing these tasks dominates the overall complexity, and the hierarchical relationship between EPs can be exploited to reduce the computational cost.

A key computational component is the soft weight calculation $\zeta(\underline{e}) = \sum_{i=1}^{N}e_i\ell_i$, which involves $\mathcal{O}(N)$ floating-point operations. By leveraging the parent-child relationship in the EP tree and maintaining auxiliary information from parent nodes, we derive an efficient recursive formulation. For a given node $\underline{e}$ with parent $\underline{e}^{(P)}$ (with $\zeta(\underline{e}^{(P)})$ and $j^{*}(\underline{e}^{(P)}) = \max_{j}\{e_{r_j}^{(P)} = 1\}$), we obtain:

\vspace{-0.3cm}
{\small
\begin{equation}\label{eq:quick_zeta}
    \zeta(\underline{e}) = \left\{\begin{array}{ll}
        \zeta(\underline{e}^{(P)}) + \ell_{r_{j^{*}+1}} - \ell_{r_{j^{*}}} & \text{if }  \underline{e} \text{ is left child} \\
        \zeta(\underline{e}^{(P)}) + \ell_{r_{j^{*}+1}}  & \text{if } \underline{e} \text{ is right child}
    \end{array}\right.,
\end{equation}
}

\noindent with corresponding depth index $j^{*}(\underline{e}) = j^{*}(\underline{e}^{(P)}) + 1$. This formulation reduces the complexity of both computing $\zeta(\cdot)$ and determining $j^{*}(\cdot)$ from $\mathcal{O}(N)$ to $\mathcal{O}(1)$.

For linear block codes, we further improve the codeword verification process through incremental computation. While standard verification using the parity-check matrix $H^{(N-K)\times N} = [\underline{h}(1),...,\underline{h}(N)]$ typically requires $N(N-K)$ operations via:
\begin{equation}
    H\cdot(\theta(y)\oplus \underline{e}) = \underline{0}\ \Leftrightarrow\  H\cdot\theta(y) = H\cdot\underline{e},
\end{equation}
we exploit the results available at the parent node to derive:

\vspace{-0.3cm}
{\small
\begin{equation}\label{eq:quick_check}
    H\cdot\underline{e} = \left\{\begin{array}{ll}
        H\cdot\underline{e}^{(P)} \oplus \underline{h}(r_{j^{*}+1}) \oplus \underline{h}(r_{j^{*}}) & \text{if } \underline{e} \text{ is left child} \\
        H\cdot\underline{e}^{(P)} \oplus \underline{h}(r_{j^{*}+1})  & \text{if } \underline{e} \text{ is right child}
    \end{array}\right..
\end{equation}
}

\noindent This transformation reduces the computational complexity from $\mathcal{O}(N(N-K))$ to $\mathcal{O}(N-K)$.

This modification requires additional space complexity to store $\{\zeta(\cdot),j^{*},H\cdot\underline{e}\}$ in exchange for significantly lower computational complexity. However, the required storage overhead is modest, as the primary memory usage is dedicated to maintaining $\mathcal{S}$.

\subsection{Early Termination Based on Code Weight}\label{subsec:4-3}

After obtaining a decoding result $\underline{e}^{*}$, the algorithm typically continues until $\zeta(\underline{e}^{*}) < \min_{\underline{e}:\underline{e} \in \mathcal{S}_{k}}\{\zeta(\underline{e})\}$ is satisfied. However, when a valid EP is discovered early with few flipped bits, we can employ the following sufficient condition for early termination:

\begin{lemma}[\cite{taipale1991improvement}]\label{lemma:1}
    If $\underline{e}$ satisfies $\theta(\underline{y})\oplus \underline{e} \in \mathcal{C}$, then a sufficient condition for $\theta(\underline{y})\oplus \underline{e}$ to be the ML decoding result is:
    \begin{equation}\label{eq:lemma1}
        \zeta(\underline{e}) = \sum_{i=1}^{N}e_i\ell_i \leq \sum_{\ell \in D(\underline{e},d_{\text{min}})} \ell \ ,
    \end{equation}
    where $d_{\text{min}}$ denotes the minimum code distance of $\mathcal{C}$, the set $D(\underline{e},d_{\text{min}})$ contains the $d_{\text{min}} - \sum_{i=1}^{N}e_i$ smallest elements from $\{\ell_{i}\mid i:e_{i} = 0\}$. If $d_{\text{min}} \leq \sum_{i=1}^{N}e_i$, we set $D(\underline{e},d_{\text{min}}) = \emptyset$.
\end{lemma}
\begin{IEEEproof}
    See \cite{taipale1991improvement} and \cite[Chapter 10.3]{Lin2004ErrorCC}.
\end{IEEEproof}

To illustrate the application of Lemma \ref{lemma:1}, consider the following example:
\begin{example}\label{ex:3}
    Consider a code with a minimum code distance $d_{\text{min}} = 4$. Assume that the received reliability values are $\underline{\ell} = (0.8, 1.2, 2.1, 3.4, 5.0, 6.0, 7.0)$. If we identify $\underline{e}^{*} = (0001000)$ as a valid EP during decoding, then:
    \begin{equation*}
        d_{\text{min}} - \sum_{i=1}^{N}e_i = 4 - 1 = 3, \ D(\underline{e},d_{\text{min}}) = \{0.8, 1.2, 2.1 \}.
    \end{equation*}
    Since $\zeta(\underline{e}^{*}) = 3.4 < 0.8 + 1.2 + 2.1 = 4.1$, we can directly declare that $(0001000)$ achieves ML decoding, without waiting for the condition $\zeta(\underline{e}^{*}) < \min_{\underline{e}:\underline{e} \in \mathcal{S}_{k}}\{\zeta(\underline{e})\}$ to be satisfied.
\end{example}

The computational complexity of applying Lemma \ref{lemma:1} is $\mathcal{O}(d_{\text{min}})$. Under high SNR conditions, valid EPs can often be identified quickly with large $d_{\text{min}} - \sum_{i=1}^{N}e_i$, allowing early termination and significantly reducing EP evaluations and decoding latency.

We note that the above early termination criterion relies on the knowledge of the minimum distance $d_{\text{min}}$ of the code. Therefore, it may not be directly applicable to codes for which $d_{\text{min}}$ is unknown or difficult to obtain. In such cases, the decoding algorithm can still operate without applying this criterion.

\subsection{Complexity Analysis of Parallel SGRAND}\label{subsec:4-4}

Based on the acceleration techniques developed in the preceding three subsections, we now present the algorithm of parallel SGRAND, as shown in Algorithm \ref{alg:extended_parallel_SGRAND}.

\begin{algorithm}[htbp]
    \caption{Parallel SGRAND}\label{alg:extended_parallel_SGRAND}
    \textbf{Inputs}: $\underline{y}$, $k_{max}$, $H$\;
    \textbf{Outputs}: $\hat{\underline{w}}$\;
    $k = 0$, $\zeta_{\text{min}}= +\infty$, $\underline{r}\gets $ reliability ranking vector, $\mathcal{S}_{0} \gets \{\underline{0}\}$, $\underline{e}^{*} \gets \emptyset$ \;
    \While{$k < k_{max}$}{
        $k \gets k + 1$\;
        
        $\mathcal{E}_{k} \gets \text{SelectNMin}(\mathcal{S}_{k-1},\zeta(\cdot),P_{k})$ \tcp*{Step 2 in Section \ref{subsec:basic-parallel-SGRAND}}
        \If{ $\sum_{\underline{e}\in \mathcal{E}_{k}}\mathbf{1}(H\cdot\theta(\underline{y}) = H\cdot\underline{e}) \geq 1$}{
            $\underline{\hat{e}} \gets \arg\min_{\underline{e} \in \mathcal{E}_{k}:\theta(\underline{y})\oplus \underline{e} \in \mathcal{C}} \zeta(\underline{e})$\;
            \If{$\zeta(\underline{\hat{e}}) < \zeta_{min}$}{
                $\underline{e}^{*} \gets \underline{\hat{e}}$, 
                $\zeta_{min} \gets \zeta(\underline{\hat{e}})$\;
                \If{$\underline{e}^{*}$ satisfies condition \eqref{eq:lemma1}}{
                    \Return $\hat{\underline{w}} = \theta(\underline{y})\oplus \underline{e}^{*}$\tcp*{Early termination (Section \ref{subsec:4-3})}
                }
                $\mathcal{S}_{k} \gets \emptyset$ \tcp*{Pruning strategy (Section \ref{subsec:4-1})}
            }
        }
        Calculate$(\zeta(\underline{e}), j^{*}, H\cdot\underline{e})$ for $\underline{e} \in \mathcal{E}_{k}^{+}$ \tcp*{Tree-based acceleration (Section \ref{subsec:4-2})}
        $\mathcal{S}_{k} \gets (\mathcal{S}_{k-1}\backslash \mathcal{E}_{k}) \cup \text{Pruning}(\mathcal{E}_{k}^{+},\zeta_{\text{min}})$ \tcp*{Pruning strategy (Section \ref{subsec:4-1})}
        \If{$\tau_{k} = \min_{\underline{e}:\underline{e} \in \mathcal{S}_{k}}\{\zeta(\underline{e})\} > \zeta_{\text{min}}$}{
            \Return $\hat{\underline{w}} = \theta(\underline{y})\oplus \underline{e}^{*}$\;
        }
    }
    \Return $\hat{\underline{w}} = 
    \begin{cases} 
        \emptyset & \text{if } \zeta_{\text{min}} = +\infty, \\
        \theta(\underline{y}) \oplus \underline{e}^* & \text{otherwise.}
    \end{cases}$
\end{algorithm}

We still utilize the min-heap data structure for maintaining $\mathcal{S}$. Table \ref{tab:parallel-complexity} provides a comparison of decoding computational complexity between serial SGRAND and parallel SGRAND. The table details the complexity of selecting $P_k$ EPs from $\mathcal{S}_{k-1}$, of completing the verification process of the $k$th round in parallel, and of updating $\mathcal{S}_{k}$. For reference, we also include the complexity required to perform $P_k$ tests using serial SGRAND.

\begin{table}[htbp]
    \centering
    \renewcommand{\arraystretch}{1.35}
    \caption{Complexity comparison between serial and parallel SGRAND implementations.}
    \label{tab:parallel-complexity}
    \begin{tabular}{|cc|cc|}
    \hline
    \multicolumn{2}{|c|}{\multirow{2}{*}{Operations}}          & \multicolumn{2}{c|}{Implementation}             \\ \cline{3-4} 
    \multicolumn{2}{|c|}{}                             & \multicolumn{1}{c|}{Serial}     &  Parallel     \\ \hline
    \multicolumn{1}{|c|}{\multirow{5}{*}{\makecell{EP\\Generation} }} & Select in heap & \multicolumn{1}{c|}{$P_k\mathcal{O}(1)$}   & $P_k\mathcal{O}(1)$    \\ \cline{2-4} 
    \multicolumn{1}{|c|}{}                        & Remove from heap  & \multicolumn{1}{c|}{$P_k\mathcal{O}(\log|\mathcal{S}_{k}|)$}   & $P_k\mathcal{O}(\log|\mathcal{S}_{k}|)$ \\ \cline{2-4} 
    \multicolumn{1}{|c|}{}                        & Generate  & \multicolumn{1}{c|}{$P_k\mathcal{O}(N)$}   & $\mathcal{O}(1)$    \\ \cline{2-4} 
    \multicolumn{1}{|c|}{}                        & Calculate $\zeta(\cdot)$& \multicolumn{1}{c|}{$P_k\mathcal{O}(N)$}   & $\mathcal{O}(1)$ \\ \cline{2-4} 
    \multicolumn{1}{|c|}{}                        & Insert to heap & \multicolumn{1}{c|}{$P_k\mathcal{O}(\log|\mathcal{S}_{k}|)$}   & $P_k\mathcal{O}(\log|\mathcal{S}_{k}|)$ \\ \hline
    \multicolumn{2}{|c|}{Codeword Test}                         & \multicolumn{1}{c|}{$P_k\mathcal{O}(N)$} &  \multicolumn{1}{c|}{$\mathcal{O}(N-K)$}       \\ \hline
    \end{tabular}
\end{table}

Heap operations exhibit similar complexity in both serial and parallel implementations. In parallel SGRAND, leveraging the hierarchical relationship between EPs and their children reduces the cost of a single computation of $\zeta(\cdot)$ from $\mathcal{O}(N)$ to $\mathcal{O}(1)$, and this operation can also be processed in parallel. The complexity of codeword verification slightly differs between implementations. Serial SGRAND checks parity equations one by one and can thus exit early when a failure occurs, while on average the complexity is of $\mathcal{O}(N)$. Parallel SGRAND uses a tree structure to accelerate verification which can be executed in parallel, resulting in a complexity of only $\mathcal{O}(N-K)$.

We remark that, while the acceleration techniques can also be applied to serial SGRAND, the primary advantage of parallel SGRAND lies in its lower latency, accomplished through concurrent computation through multiple threads or matrix operations.

\section{ML Enhancement of ORBGRAND}\label{sec:orbgrand}


ORBGRAND is known to admit highly efficient implementations, and we further observe that it can be naturally interpreted within the same EP-tree framework introduced earlier. This observation motivates a hybrid decoding strategy that combines ORBGRAND with the proposed tree-based approach. In the resulting scheme, the majority of EPs are generated and verified through ORBGRAND, while the EP tree-based mechanisms are invoked selectively to ensure ML optimality. By concentrating tree-based operations on critical parts of the decoding process, the proposed approach achieves improved decoding efficiency within a unified framework.

This section proceeds as follows. We first establish an EP-tree-based interpretation of ORBGRAND (Section~\ref{subsec:5-1}). We then present the resulting hybrid enhanced ORBGRAND algorithm (Section~\ref{subsec:5-2}). Finally, we analyze the computational complexity of the proposed scheme (Section~\ref{subsec:5-3}).

\subsection{ORBGRAND and EP Tree Representation}\label{subsec:5-1}

While SGRAND attains ML decoding performance, its online EP generation incurs substantial runtime overhead. To address this limitation, \cite{duffy2022ordered} introduced ORBGRAND, which generates EPs based solely on the ranking of the received channel reliabilities, rather than their exact values.

Subsequently, several ORB-type GRAND algorithms have been proposed~\cite{liu2022orbgrand,wan2024approaching}. Although EPs can be generated either on-the-fly during decoding~\cite{duffy2022ordered,yuan2023guessing} or pre-computed and stored~\cite{9715725}, from an analytical perspective these algorithms can be equivalently viewed as operating on a fixed EP set
\[
\tilde{\mathcal{E}}_{\mathrm{ORB}} = \{\tilde{\underline{e}}(1), \ldots, \tilde{\underline{e}}(T)\},
\]
where, for each $t$, the runtime EP is obtained by applying the same $\underline{r}$-induced coordinate permutation to the $\tilde{\underline{e}}(t)$.

Under this abstraction, ORB-type GRAND decoding can be described by two conceptual phases:
\begin{itemize}
    \item \emph{EP definition phase}: Define a reliability-ordered EP set $\tilde{\mathcal{E}}_{\mathrm{ORB}}$, independent of the received vector realization.
    \item \emph{Runtime phase}: For each received vector $\underline{y}$, apply the $\underline{r}$-induced coordinate permutation to each EP vector to generate the actual testing patterns.
\end{itemize}

Different ORB-type GRAND algorithms have different $\tilde{\mathcal{E}}_{\text{ORB}}$ sets and testing schedules.

\begin{definition}[$\gamma$-based $\tilde{\mathcal{E}}_{\text{ORB}}$]\label{def:gamma-grand}
    Given a non-decreasing weight vector $\underline{\gamma}$ of length $N$, i.e., satisfying $\gamma_i \leq \gamma_j$ for any $1 \leq i < j \leq N$, the $\gamma$-based $\tilde{\mathcal{E}}_{\text{ORB}}$ comprises of the first $T$ binary vectors $\Tilde{\underline{e}}(t)$ in $\mathbb{F}_2^N$ that satisfy the ordering constraint:
    \begin{equation}\label{eq:orb_tilde}
        \sum_{i = 1}^N \gamma_i \Tilde{e}_i(t) \leq \sum_{i = 1}^N \gamma_i \Tilde{e}_i(t'), \quad \forall \ t \leq t'.
    \end{equation}
\end{definition}

During runtime, each EP $\Tilde{\underline{e}}(t)$ undergoes position permutation according to the reliability ranking vector $\underline{r}$, yielding the actual EP $\underline{e}(t)$ for testing. The algorithm thus tests EPs from the sequence $\mathcal{E}_{\text{ORB}}=\{\underline{e}(1),...,\underline{e}(T)\}$ in order, satisfying:
\begin{equation}\label{eq:orb}
    \sum_{i=1}^{N}\gamma_{i} e_{r_i}(t) \leq \sum_{i=1}^{N}\gamma_{i} e_{r_i}(t'), \quad \forall \ t \leq t'. 
\end{equation}

For the special case where $\gamma_i = i$, $\forall i = 1, \ldots, N$, we have the original ORBGRAND algorithm \cite{duffy2022ordered}.

While ORB-type GRAND algorithms reduce the runtime overhead of EP generation and hence enables efficient parallelization, they cannot guarantee ML decoding due to their reliance on reliability ranking rather than exact channel reliability values. This drawback motivates us to investigate the possibility of enhance ORB-type GRAND algorithms with the EP tree representation developed in the previous sections.

The following proposition establishes the theoretical foundation for our algorithm design:

\begin{proposition}\label{prop:sub_tree}
    Given $\mathcal{E}_{\text{ORB}}$, for any $T_0 \leq T$, the first $T_0$ EPs of $\mathcal{E}_{\text{ORB}}$, denoted by $\mathcal{E}_0$, form a subtree of the complete EP tree that includes the root node.
\end{proposition}
\begin{IEEEproof}
    Consider the complete EP tree containing all $2^{N}$ possible EPs. To show that $\mathcal{E}_0$ forms a subtree, it suffices to prove that for any $\underline{e}\in\mathcal{E}_0$, its parent node also belongs to $\mathcal{E}_0$.

    Recall that EPs in $\mathcal{E}_{\mathrm{ORB}}$ (and hence in $\mathcal{E}_0$) satisfy~\eqref{eq:orb}. 
    Define the soft weight $\gamma(\underline{e},\underline{r}) = \sum_{i=1}^{N}\gamma_{i} e_{r_i}$. By construction, EPs are visited in increasing order of $\gamma(\underline{e},\underline{r})$. Therefore, it suffices to show that the parent node $\underline{e}^{(P)}$ of any $\underline{e}\in\mathcal{E}_0$ satisfies $\gamma(\underline{e}^{(P)},\underline{r}) < \gamma(\underline{e},\underline{r})$. If $\underline{e}$ is the right child:
    \begin{align}
        & \sum_{i=1}^{N}\gamma_{i} e_{r_i} - \sum_{i=1}^{N}\gamma_{i} e^{(P)}_{r_i}  \notag \\
        = &\sum_{i\neq j^{*}}\gamma_{i} e_{r_i} + \gamma_{j^{*}} e_{r_{j^{*}}} - \left(\sum_{i\neq j^{*}}\gamma_{i} e^{(P)}_{r_i} + \gamma_{j^{*}} e^{(P)}_{r_{j^{*}}}\right) \label{eq:parent-1}\\
        = &\gamma_{j^{*}} > 0.\label{eq:parent-2}
    \end{align}
    In \eqref{eq:parent-1}, $j^{*}= \max_{j}\{e_{r_j}(t) \neq 0\}$. Based on the relationship between a parent node and its right child in Definition \ref{def:tree}, we observe that all positions except for $r_{j^{*}}$ remain identical, with $e_{r_{j^{*}}} = 1$ and $e^{(P)}_{r_{j^{*}}} = 0$. 
    
    If $\underline{e}$ is the left child, then by Definition~\ref{def:tree} the parent and child differ only at two adjacent ranked positions, namely $(r_{j^{*}-1}, r_{j^{*}})$. Hence,
    \begin{equation}\label{eq:parent-3}
    \sum_{i=1}^{N}\gamma_{i} e_{r_i}-\sum_{i=1}^{N}\gamma_{i} e^{(P)}_{r_i}
    = \gamma_{j^{*}}-\gamma_{j^{*}-1} > 0,
    \end{equation}
    where the last inequality follows from the monotone ordering of $\{\gamma_i\}$ induced by the reliability ranking.

    Combining \eqref{eq:parent-2} and \eqref{eq:parent-3}, we establish that the parent node of any $\underline{e} \in \mathcal{E}_0$ also belongs to $\mathcal{E}_0$. By induction, we have that all ancestors of $\underline{e}$ belong to $\mathcal{E}_0$. Therefore, $\mathcal{E}_0$ constitutes a subtree of the EP tree that contains the root node.
\end{IEEEproof}
\begin{remark}
    There also exist several ORB-type GRAND algorithms beyond $\gamma$-based $\tilde{\mathcal{E}}_{\text{ORB}}$, including iLWO GRAND \cite{condo2021iLWO} and RS-ORBGRAND \cite{wan2024approaching}. In general, EP sequences generated by those algorithms still satisfy Proposition \ref{prop:sub_tree}. This is because the EP tree structure inherently satisfies the property that the $\zeta(\cdot)$ of a parent node is always smaller than that of its children, and ORB-type GRAND algorithms generally follow the principle of visiting parent nodes before their child nodes.
\end{remark}

\subsection{Hybrid Enhanced ORBGRAND}\label{subsec:5-2}

Building upon the preceding analysis, we develop a hybrid decoding approach that leverages ORB-type GRAND algorithms for initial decoding, followed by the parallel SGRAND decoding in the previous section, thereby achieving improved parallelism and reduced computational complexity. The algorithm is described in Algorithm \ref{alg:ORB_with_SGRAND}.
\begin{algorithm}[htbp]
    \caption{Hybrid Enhanced ORBGRAND}\label{alg:ORB_with_SGRAND}
    \textbf{Inputs}: $\underline{y}$, $k_{\text{max}}$, $H$, $\tilde{\mathcal{E}}_{\text{ORB}}$\;
    \textbf{Outputs}: $\hat{\underline{w}}$\;
    $k = 0$, $\mathcal{E}_0 \gets \{\underline{0}\}$, $\underline{e}^{*} \gets \emptyset$, $\zeta_{\text{min}}= +\infty$, $\underline{r}\gets $ reliability ranking vector\;
    $\mathcal{E}_{\text{ORB}} \gets \text{Permute}(\tilde{\mathcal{E}}_{\text{ORB}}, \underline{r})$ \tcp*{See \eqref{eq:orb}}
    \For{$\underline{e} \in \mathcal{E}_{\text{ORB}}$}{
        \If{$H\cdot\theta(\underline{y}) = H\cdot\underline{e}$}{
            $\underline{e}^{*} \gets \underline{e}$, $\zeta_{min} \gets \zeta(\underline{e})$\;
            \textbf{break}\;
        }
    }
    $\mathcal{E}_0 \gets $ Tested EPs \;
    $\mathcal{S}_0\gets $ Envelope of $ \mathcal{E}_0$ \;
    \Return $\hat{\underline{w}} \gets \text{Parallel SGRAND}(\underline{e}^{*}, \zeta_{\text{min}}, \mathcal{S}_0)$ \tcp*{Call Algorithm \ref{alg:extended_parallel_SGRAND} with initial values of $\underline{e}^{*}, \zeta_{\text{min}}, \mathcal{S}_0$ in Line 3 replaced}
\end{algorithm}

In Algorithm \ref{alg:ORB_with_SGRAND}, after computing $\underline{r}$, we apply it to the pre-generated EP set $\tilde{\mathcal{E}}_{\text{ORB}}$ to obtain $\mathcal{E}_{\text{ORB}}$. The algorithm then performs the following two main operations, as illustrated in Figure \ref{fig:orb_ml_method}:
\begin{enumerate}
    \item Test the elements in $\mathcal{E}_{\text{ORB}}$. If a valid EP is found, terminate decoding and record the tested EPs.
    \item Denote the envelope of the tested EPs as $\mathcal{S}_{0}$, and use it as input to Algorithm \ref{alg:extended_parallel_SGRAND} to continue decoding.
\end{enumerate}
The envelope of $\mathcal{E}_0$ refers to the nodes in the EP tree (excluding $\mathcal{E}_0$) that have a distance one to at least one leaf node of $\mathcal{E}_0$.

\begin{figure}[tbp]
\centering
\begin{tikzpicture}[
    scale=0.95,
    set/.style={
        circle,
        draw,
        minimum size=1.2cm,
        align=center,
        font=\small,
        fill=white
    },
    checker/.style={
        rectangle,
        draw,
        minimum width=2.5cm,
        minimum height=0.4cm,
        align=center,
        fill=blue!8,
        inner sep=0.5pt
    },
    checkerGroup/.style={
        rectangle,
        draw,
        rounded corners=5pt,
        inner xsep=4pt,
        inner ysep=4pt,
        fill=blue!3,
        dashed
    },
    rect/.style={
        rectangle,
        draw,
        rounded corners=3pt,
        minimum width=1.9cm,
        minimum height=0.85cm,
        align=center,
        font=\small,
        fill=white
    },
    arrow/.style={
        ->, 
        >=stealth, 
        thick,
        shorten >=2pt,
        shorten <=2pt
    },
    time/.style={
        font=\footnotesize,
        above right=-0.3cm and 0.1cm,
        text=gray
    },
    scale=0.88,
    transform shape
]

\node[set, fill=green!15] (e0) at (0, -4.7) {$\mathcal{E}_{\text{ORB}}$};

\foreach \i in {0,...,3} {
    \node[checker] (checker0-\i) at (0, -1.6 - \i*0.5) {Tester \pgfmathparse{int(\i+1)}\pgfmathresult};
}

\begin{scope}[on background layer]
    \node[checkerGroup, fit=(checker0-0)(checker0-3)(checker0-3.south east)] {};
\end{scope}

\node[set, fill=green!15] (ex) at (-1.25, 0) {Tested \\ $\mathcal{E}_0$};

\draw[arrow] (e0) -- (0, -3.5);
\draw[arrow] (0, -1.1) -- (ex);

\node[set, fill=yellow!20] (s0) at (1.5, 0) {$\mathcal{S}_0$};
\draw[dashed, gray] (2.15, -2.5) -- (2.15, 1) node[time] {$k=1$};

\node[set] (e1) at (4, 0) {$\mathcal{S}_0 \backslash\mathcal{E}_{1}$};
\node[set, fill=green!15] (e1d) at (3, -1.7) {$\mathcal{E}_{1}$};

\foreach \i in {0,...,3} {
    \node[checker] (checker1-\i) at (4, -3.4 - \i*0.5) {Tester \pgfmathparse{int(\i+1)}\pgfmathresult};
}

\begin{scope}[on background layer]
    \node[checkerGroup, fit=(checker1-0)(checker1-3)(checker1-3.south east)] {};
\end{scope}

\node[set] (e1child) at (5.0, -1.7) {$\mathcal{E}_{1}^{+}$};

\draw[arrow] (ex) -- node[above, pos=0.4] {envelope}(s0);
\draw[arrow] (s0) -- (e1);
\draw[arrow] (s0) -- node[right, pos=0.4] {top $P_1$} (e1d);
\draw[arrow] (e1d) -- (3, -3);
\draw[arrow] (5.0, -3) -- (e1child);

\node[set, fill=yellow!20] (s1) at (6.5, 0) {$\mathcal{S}_1$};
\draw[arrow] (e1) -- (s1);
\draw[arrow] (e1child) -- (s1);
\node at (7.5, 0) {$\cdots$};

\end{tikzpicture}
\caption{Schematic diagram of the decoding process of Algorithm \ref{alg:ORB_with_SGRAND}.}
\label{fig:orb_ml_method}
\end{figure}

\begin{example}\label{ex:4}
We still work with the case in Example \ref{ex:tree_sgrand}, with $\underline{\ell} = (0.8, 1.2, 2.1, 3.4)$ and $\underline{r}=\{1, 2, 3, 4\}$. Letting
$$\tilde{\mathcal{E}}_{\text{ORB}}=\{(0000), (1000), (0100), (0010), (1100), (1010)\},$$
then under $\underline{r}$, by permutation, we obtain $\mathcal{E}_{\text{ORB}} = \tilde{\mathcal{E}}_{\text{ORB}}$.
If we test the first 4 EPs therein and find that the $4$-th one, $(0010)$, leads to a valid codeword, then the 4 tested EPs constitute $\mathcal{E}_0$, and the envelope $\mathcal{S}_0$ is
$$\mathcal{S}_{0}=\{(0001), (0011), (0110), (1100)\}.$$
The process is illustrated in Figure \ref{fig:ex_tree_quick}, in which the tested EPs are marked in black, their envelope $\mathcal{S}_0$ in red, and the remaining EPs in blue. Then we create a min-heap based on $\mathcal{S}_0$ and continue decoding using Algorithm \ref{alg:extended_parallel_SGRAND}.

\begin{figure}[htbp]
    \centering
    \resizebox{0.48\textwidth}{!}{%
    \begin{tikzpicture}
    \tikzset{level distance=19pt, sibling distance=1pt}
    \Tree
    [.0000
        [.1000
            [.0100
                [.0010
                    [.\textcolor{red}{0001} ]
                    [.\textcolor{red}{0011} ]
                ]
                [.\textcolor{red}{0110} 
                    \edge[dashed];[.\textcolor{gray}{0101} ]
                    \edge[dashed];[.\textcolor{gray}{0111} ]
                ]
            ]
            [.\textcolor{red}{1100}
                \edge[dashed];[.\textcolor{gray}{1010} 
                    \edge[dashed];[.\textcolor{gray}{1001} ]
                    \edge[dashed];[.\textcolor{gray}{1011} ]
                ]
                \edge[dashed];[.\textcolor{gray}{1110} 
                    \edge[dashed];[.\textcolor{gray}{1101} ]
                    \edge[dashed];[.\textcolor{gray}{1111} ]
                ]
            ]
        ]
    ]
    \end{tikzpicture}
    }
    \caption{Example \ref{ex:4}: Elements of $\mathcal{S}_{0}$ are marked in red for subsequent parallel SGRAND decoding.}
    \label{fig:ex_tree_quick}
\end{figure}
\end{example}

In the following theorem, we show that this algorithm achieves ML decoding.

\begin{theorem}\label{th:2}
    Algorithm \ref{alg:ORB_with_SGRAND} tests each EP at most once, and when the EP tree is complete with all the $2^N$ possible EPs, it achieves ML decoding.
\end{theorem}
\begin{IEEEproof}
    According to Proposition \ref{prop:sub_tree}, the tested EPs (i.e., $\mathcal{E}_0$) form a subtree of the EP tree that contains the root node, and each EP is tested only once. Due to this fact, the nodes in the envelope $\mathcal{S}_0$ are neither parent or child of any other. We can hence use Theorem \ref{th:1} to arrive at the conclusion that the descendants of $\mathcal{S}_0$ will only be visited at most once. Consequently, we have
    \begin{align}
        &\mathcal{E}_0 \cup (\mathcal{S}_0 \cup \text{descendants of }\mathcal{S}_0) = 2^N, \\
        &\mathcal{E}_0 \cap (\mathcal{S}_0 \cup \text{descendants of }\mathcal{S}_0) = \emptyset,
    \end{align}
    thereby guaranteeing non-repetitive traversal of any element of the EP tree.
    
    With $\tau_0 = \min_{\underline{e}:\underline{e} \in \mathcal{S}_{0}}\{\zeta(\underline{e})\}$, recalling the proof of Theorem \ref{th:1}, we still have $\tau_0 \leq \tau_1 \leq ...$, and continuously updating the locally optimal EPs satisfies the descending order of $\zeta(\cdot)$. This ensures that the EP identified by the algorithm is optimal achieving ML decoding.
\end{IEEEproof}

Invoking the ORB-type GRAND algorithm to obtain $\mathcal{E}_0$ greatly saves the runtime overhead of generating EPs online. On the other hand, calculating the envelope $\mathcal{S}_0$ and maintaining the min-heap incur certain additional computational overhead. In the next subsection, we will systematically analyze the complexity of the algorithms proposed in this paper.

\begin{table*}[htbp]
    \scriptsize
    \setlength{\tabcolsep}{4.5pt}
    \centering
    \renewcommand{\arraystretch}{1.7}
    \caption{Analytical Computational and Space Complexity Breakdown of GRAND Algorithm Variants. The sorting cost is reported as an algorithm-level upper bound; see Remark~\ref{rem:sorting} for hardware-level alternatives.}
    \begin{tabular}{|l|c|c|c|c|c|c|}
    \hline
    & \multicolumn{2}{c|}{\textbf{Type}} 
    & \textbf{SGRAND} 
    & \textbf{ORB-type GRAND} 
    & \textbf{Parallel SGRAND} 
    & \textbf{Hybrid enhanced ORBGRAND} \\ \hline

    & \multicolumn{2}{c|}{\textbf{Algorithm}} 
    & Alg.~\ref{alg:SGRAND_tree}~\cite{solomon2020soft} 
    & \cite{duffy2022ordered},~\cite{wan2024approaching} 
    & Alg.~\ref{alg:extended_parallel_SGRAND} 
    & Alg.~\ref{alg:ORB_with_SGRAND} \\ \hline
    \hline

    \multirow{5}{*}{\rotatebox[origin=c]{90}{\textbf{Computational}}} 
    & \multicolumn{2}{c|}{Sorting} 
    & $\mathcal{O}(N \log N)$ 
    & $\mathcal{O}(N \log N)$ 
    & $\mathcal{O}(N \log N)$ 
    & $\mathcal{O}(N \log N)$ \\ \cline{2-7}

    & \multicolumn{2}{c|}{Codeword Tester (XOR)} 
    & $T_{\mathrm{SG}} \cdot c \cdot N$ 
    & $T_{\mathrm{OG}} \cdot N (N-K)/P$ 
    & $T_{\mathrm{PSG}} \cdot (N-K)/P$ 
    & $[T_{\mathrm{OG}} \cdot N (N-K) + (T_{\mathrm{HOG}}-T_{\mathrm{OG}})\cdot (N-K)]/P$ \\ \cline{2-7}

    & \multirow{3}{*}{\makecell{EP\\Generation}} 
    & XOR 
    & $T_{\mathrm{SG}} \cdot \mathcal{O}(N)$ 
    & \multirow{3}{*}{$\mathcal{O}(1)$} 
    & $T_{\mathrm{PSG}} \cdot \mathcal{O}(N-K)/P$ 
    & $\mathcal{O}(1) + (T_{\mathrm{HOG}}-T_{\mathrm{OG}})\cdot \mathcal{O}(1)/P$ \\ \cline{3-4} \cline{6-7}

    &  
    & Real Addition 
    & $T_{\mathrm{SG}} \cdot \mathcal{O}(N)$ 
    &  
    & $T_{\mathrm{PSG}} \cdot \mathcal{O}(1)/P$ 
    & $T_{\mathrm{OG}} \cdot \mathcal{O}(N) + (T_{\mathrm{HOG}}-T_{\mathrm{OG}})\cdot \mathcal{O}(1)/P$ \\ \cline{3-4} \cline{6-7}

    &  
    & Comparison 
    & $\mathcal{O}(T_{\mathrm{SG}} \log T_{\mathrm{SG}})$ 
    &  
    & $\mathcal{O}(T_{\mathrm{PSG}} \log T_{\mathrm{PSG}})$ 
    & $T_{\mathrm{OG}} \cdot \mathcal{O}(1) + \mathcal{O}((T_{\mathrm{HOG}}-T_{\mathrm{OG}})\log (T_{\mathrm{HOG}}-T_{\mathrm{OG}}))$ \\ \hline
    \hline

    \multirow{4}{*}{\rotatebox[origin=c]{90}{\textbf{Space}}} 
    & \multirow{2}{*}{Codeword} 
    & Bits 
    & $N$ & $N$ & $N$ & $N$ \\ \cline{3-7}

    &  
    & Floats 
    & $N$ & $N$ & $N$ & $N$ \\ \cline{2-7}

    & \multirow{2}{*}{EPs} 
    & Bits 
    & $T_{\mathrm{SG}} \cdot N$ 
    & $/$ 
    & $\leq T_{\mathrm{PSG}} \cdot (2N-K)$ 
    & $\leq (T_{\mathrm{HOG}}-T_{\mathrm{OG}})\cdot (2N-K)$ \\ \cline{3-7}

    &  
    & Floats 
    & $T_{\mathrm{SG}}$ 
    & $/$ 
    & $\leq T_{\mathrm{PSG}} \cdot 2$ 
    & $\leq T_{\mathrm{HOG}} \cdot 2$ \\ \hline
    \end{tabular}
    \label{tab:grand_complexity}
\end{table*}

\subsection{Complexity Analysis}\label{subsec:5-3}

Table \ref{tab:grand_complexity} presents a comprehensive complexity comparison among the proposed algorithms. We denote the number of tests for SGRAND, ORBGRAND, parallel SGRAND, and hybrid enhanced ORBGRAND as $T_{\mathrm{SG}}$, $T_{\mathrm{OG}}$, $T_{\mathrm{PSG}}$, and $T_{\mathrm{HOG}}$, respectively. These numbers can be estimated empirically via Monte Carlo simulation runs. We have $T_{\mathrm{HOG}} \geq T_{\mathrm{OG}}$ since hybrid enhanced ORBGRAND includes an initial ORBGRAND phase before the subsequent parallel SGRAND phase.

The computational complexity results presented in Table \ref{tab:grand_complexity} already account for parallelization. That is, the computational complexity of codeword verification is scaled by a parallelization factor $1/P$, where $P$ accounts for the average degree of parallelism.\footnote{For simplicity, in our numerical simulation reported in the next section, we use the same value of $P_k = P$ for all $k$ when implementing parallel SGRAND. It is certainly possible to use different values of $P_k$ in different rounds to further optimize the speedup, and we leave this as a possible future research topic.}

For codeword verification, SGRAND performs $T_{\mathrm{SG}} c N$ tests with early termination, where $c$ denotes the average fraction of verified symbols upon termination. ORBGRAND exploits parallel computing to achieve reduced complexity of $T_{\mathrm{OG}} \cdot N(N-K)/P$ through parallelization factor $P$. Parallel SGRAND further optimizes this process, reducing complexity to $T_{\mathrm{PSG}} \cdot (N-K)/P$ by leveraging hierarchical EP relationship as described in Section \ref{subsec:4-2}. Hybrid enhanced ORBGRAND combines both phases: initially verifying $T_{\mathrm{OG}}$ EPs using ORBGRAND, then processing the remaining $T_{\mathrm{HOG}}-T_{\mathrm{OG}}$ EPs through parallel SGRAND.

The complexity of generating EPs differs greatly. SGRAND requires $T_{\mathrm{SG}} \cdot \mathcal{O}(N)$ operations for both XOR and additions, and EP tree maintenance incurs $\mathcal{O}(T_{\mathrm{SG}} \log T_{\mathrm{SG}})$ complexity. Parallel SGRAND optimizes computation through exploitation of the tree structure, reducing calculations to $\mathcal{O}(1)$ and can be operate in parallel, as described in Section \ref{subsec:4-2}. ORBGRAND achieves superior efficiency by directly generating EPs from $\tilde{\mathcal{E}}_{\text{ORB}}$ and the reliability ranking vector $\underline{r}$, thereby incurring negligible runtime overhead for EP generation.\footnote{In practice, EPs in ORBGRAND can be either pre-stored~\cite{9715725} or generated efficiently on the fly~\cite{duffy2022ordered,yuan2023guessing}.} After completing the phase of ORBGRAND, hybrid enhanced ORBGRAND requires $T_{\mathrm{OG}}\cdot\mathcal{O}(N)$ complexity to construct the envelope $\mathcal{S}_0$ and $T_{\mathrm{OG}} \cdot \mathcal{O}(1)$ complexity to build a min-heap for completing the subsequent parallel SGRAND phase \cite[Chapter 6]{cormen2022introduction}.

In terms of space complexity, ORBGRAND stores $\tilde{\mathcal{E}}_{\text{ORB}}$ but requires no additional runtime space. SGRAND dynamically maintains up to $T_{\mathrm{SG}}$ EPs with their soft weights. Parallel SGRAND requires additional storage for syndrome values and position indices, though pruning reduces actual EP storage needed. Hybrid enhanced ORBGRAND optimizes storage by generating only the final $T_{\mathrm{HOG}}-T_{\mathrm{OG}}$ EPs in runtime.

\begin{remark}\label{rem:sorting}
The $\mathcal{O}(N \log N)$ sorting cost in Table~\ref{tab:grand_complexity} is an algorithm-level upper bound that assumes a fully sorted reliability list prior to EP querying. Recently reported GRAND ASICs instead pipeline a serial sorter with codeword querying, producing the least reliable bits incrementally on demand and thereby reducing the sorting cost to $\mathcal{O}(N)$~\cite{riaz2023subpj,riaz2025subpj,kizilates2025lowlatency}. Sorting in such designs runs in parallel with querying and can terminate once a valid codeword is identified, further lowering the effective sorting overhead.
\end{remark}

\section{Simulation Results}\label{sec:simulation}

This section presents simulation results for evaluating the proposed algorithms.
Section \ref{subsec:6-1} compares array-based SGRAND, heap-based SGRAND, and ORBGRAND as a baseline.
Section \ref{subsec:6-2} evaluates the proposed parallel SGRAND and hybrid enhanced ORBGRAND in terms of complexity reduction.
Section \ref{subsec:6-3} provides ablation studies on key acceleration techniques.

Unless otherwise specified, simulations are conducted on the BCH(127,106) code over the AWGN channel \cite{Lin2004ErrorCC,bose1960class}, with additional results in Section \ref{subsec:6-4} validating generality across different codebooks and code rates.

\subsection{Heap-based SGRAND}\label{subsec:6-1}

\begin{figure*}[htbp]
    \centering
    \includegraphics[width=0.99\textwidth]{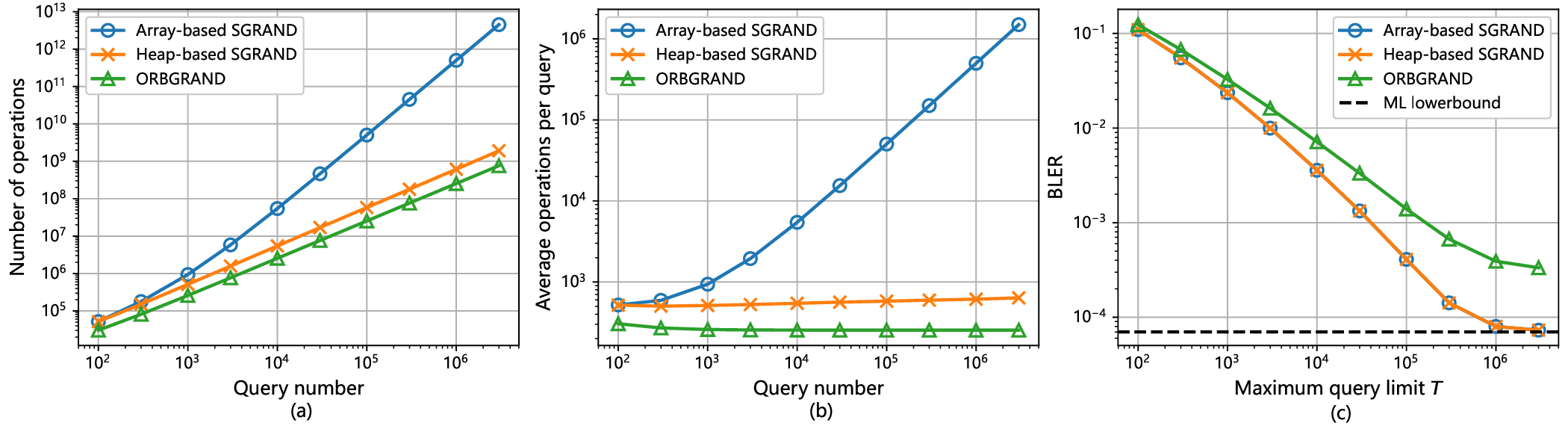}
    \caption{Performance comparison of array-based SGRAND, heap-based SGRAND, and ORBGRAND: (a) Total computational complexity $\Phi_{\mathrm{tot}}$; (b) Average computational complexity per codeword test; (c) Decoding performance.}
    \label{fig:Aim1}
\end{figure*}

This subsection evaluates the computational complexity and decoding performance of heap-based SGRAND by comparison with array-based SGRAND and ORBGRAND, using the BCH (127,106) code at $E_b/N_0 = 5$ dB. All algorithms have the same codeword testing complexity $f(N,K)$: test parities sequentially and stop on the first failure.

The reported complexity results are obtained by explicitly counting the number of operations during Monte Carlo decoding simulations. The analytical expressions in \eqref{eq:complexity_array_sum} and \eqref{eq:complexity_heap_sum} provide the underlying complexity characterization. Following commonly used practice in algorithm-level complexity analysis (e.g., \cite{yuan2023guessing}), different operations are normalized using XOR-equivalent counts: one comparison is counted as $6$ XOR operations, and one real-valued addition is counted as $8$ XOR operations. The sorting cost is evaluated as $6N\log_2 N$, accounting for the comparison operations required to order $N$ elements. Unless otherwise stated, this XOR-equivalent normalization is consistently adopted throughout the remainder of this paper for all complexity evaluations.

Figure~\ref{fig:Aim1}(a) illustrates the variation of the total computational complexity $\Phi_{\mathrm{tot}}$ with respect to the query index $t$. ORBGRAND exhibits a strictly linear growth trend. Although heap-based SGRAND contains an $\mathcal{O}(t\log t)$ term in \eqref{eq:complexity_heap_sum}, its overall complexity still grows nearly linearly with $t$. In contrast, array-based SGRAND exhibits a pronounced quadratic growth behavior, consistent with the analysis in \eqref{eq:complexity_array_sum}. Figure~\ref{fig:Aim1}(b) further compares the average complexity per codeword test, showing that both heap-based SGRAND and ORBGRAND maintain approximately constant overhead across different query limits.

Figure~\ref{fig:Aim1}(c) compares the error-correction performance under different maximum query budgets $T$. As $T$ increases, both implementations of SGRAND achieve identical ML decoding performance, whereas ORBGRAND exhibits a slight performance degradation due to its reliance on reliability ranking rather than exact reliability values.

Based on these observations, heap-based SGRAND is adopted as the baseline for subsequent evaluations.

\subsection{Parallel SGRAND and Hybrid Enhanced ORBGRAND}\label{subsec:6-2}

\begin{table*}[htbp]
    \scriptsize
    \setlength{\tabcolsep}{5.3pt}
    \centering
    \renewcommand{\arraystretch}{1.35}
    \caption{Instantiated Complexity Scores for GRAND Variants on BCH(127,106) at $E_b/N_0 = 5$ dB Under Different Parallelism Degrees.}
    \begin{tabular}{|l|c|c|c|c|c|c|c|c|c|c|}
    \hline
    & \multicolumn{2}{c|}{\textbf{Type}} 
    & \textbf{SGRAND} 
    & \textbf{RS-ORBGRAND} 
    & \multicolumn{5}{c|}{\textbf{Parallel SGRAND}} 
    & \textbf{Hybrid GRAND} \\ \hline

    & \multicolumn{2}{c|}{\textbf{Algorithm}} 
    & Alg.~\ref{alg:SGRAND_tree}~\cite{solomon2020soft} 
    & \cite{duffy2022ordered},~\cite{wan2024approaching} 
    & \multicolumn{5}{c|}{Alg.~\ref{alg:extended_parallel_SGRAND}} 
    & Alg.~\ref{alg:ORB_with_SGRAND} \\ \hline

    & \multicolumn{2}{c|}{\textbf{Parallelization $P$}} 
    & --- & $32$ 
    & $4$ & $8$ & $16$ & $32$ & $64$
    & \multicolumn{1}{c|}{$32$} \\ \hline

    & \multicolumn{2}{c|}{\textbf{Average Queries}} 
    & $272$ & $322$ 
    & $276$ & $288$ & $294$ & $312$ & $346$
    & \multicolumn{1}{c|}{$372$} \\ \hline
    \hline

    \multirow{8}{*}{\rotatebox[origin=c]{90}{\textbf{Time Complexity}}} 
    & \multicolumn{2}{c|}{Sorting (5325)} 
    & $1$ 
    & $1$ 
    & $1$ & $1$ & $1$ & $1$ & $1$
    & $1$ \\ \cline{2-11}

    & \multicolumn{2}{c|}{Codeword Tester (XOR)} 
    & $71120$ 
    & $858774$ 
    & $17388$ & $18018$ & $18522$ & $19656$ & $21798$
    & $858774$ \\ \cline{2-11}

    & \multirow{3}{*}{\makecell{EP\\Generation}} 
    & XOR (1)
    & $28576$ 
    & \multirow{3}{*}{---} 
    & $552$ & $572$ & $588$ & $624$ & $692$
    & $27547$ \\ \cline{3-4} \cline{6-11}

    &  & Real Addition (8) 
    & $2144.6$ 
    &  
    & $552$ & $572$ & $588$ & $624$ & $692$
    & $1449$ \\ \cline{3-4} \cline{6-11}

    &  & Comparison (6) 
    & $\textit{\textbf{5403.5}}$ 
    &  
    & $\textit{\textbf{5048}}\!+\!552$ & $\textit{\textbf{5264}}\!+\!572$ & $\textit{\textbf{5343}}\!+\!588$ & $\textit{\textbf{5469}}\!+\!624$ & $\textit{\textbf{5817}}\!+\!692$
    & $\textit{\textbf{280}}\!+\!550$ \\ \cline{2-11}

    & \multicolumn{2}{c|}{\textbf{Total Complexity Score ($\Phi_{\text{tot}}^{(\cdot)})$}} 
    & $154598$ & --- 
    & $61281$ & $63693$ & $64725$ & $67155$ & $72405$
    & $908218$ \\ \cline{2-11}

    & \multicolumn{2}{c|}{\textbf{Parallel Complexity Score ($\Phi_{\text{par}}^{(\cdot)})$}} 
    & --- & $32162$ 
    & $42030$ & $40257$ & $39092$ & $39045$ & $40730$
    & $36737$ \\ \cline{2-11}

    & \multicolumn{2}{c|}{\textbf{Speed Up ($\eta^{(\cdot)}$)}} 
    & $\times 1.00$ & $\times 4.81$ 
    & $\times 3.68$ & $\times 3.84$ & $\times 3.95$ & $\times 3.96$ & $\times 3.80$
    & $\times 4.21$ \\ \hline
    \hline

    \multirow{5}{*}{\rotatebox[origin=c]{90}{\textbf{Space}}} 
    & \multirow{2}{*}{Codeword} & Bits (1)
    & $127$ & $127$
    & $127$ & $127$ & $127$ & $127$ & $127$ 
    & $127$ \\ \cline{3-11}

    & & Floats (8)
    & $127$ & $127$
    & $127$ & $127$ & $127$ & $127$ & $127$ 
    & $127$ \\ \cline{2-11}

    & \multirow{2}{*}{EPs} & Bits (1)
    & $34544$ & $/$ 
    & $40700$ & $41400$ & $42476$ & $43512$ & $45140$ 
    & $7140$ \\ \cline{3-11}

    & & Floats (8) 
    & $272$ & $/$ 
    & $550$ & $560$ & $574$ & $588$ & $610$ 
    & $96$ \\ \cline{2-11}

    & \multicolumn{2}{c|}{\textbf{Total Complexity Score}} 
    & $37863$ & $1143$ 
    & $46243$ & $47023$ & $48211$ & $49359$ & $51163$
    & $9051$ \\ \hline
    \end{tabular}
    \label{tab:grand_complexity2}
\end{table*}

This subsection evaluates the implementation-level computational complexity and decoding performance of the proposed parallel SGRAND and hybrid enhanced ORBGRAND algorithms. Building on the complexity framework established in Section~\ref{subsec:5-3}, we explicitly quantify the trade-off between decoding performance and computational complexity under different degrees of parallelism.

We denote by $\Phi_{\text{tot}}^{(\cdot)}$ the total computational complexity of a decoding algorithm.
As formalized in \eqref{eq:Phi_tot}, this total complexity can be decomposed into a non-parallelizable component and a parallelizable component, where the former mainly arises from heap-related operations. 
For a given parallelism degree $P$, the effective computational complexity $\Phi_{\text{eff}}^{(\cdot)}$ under parallel execution is defined in \eqref{eq:Phi_eff}, in which only the parallelizable portion is reduced by a factor of $P$, allowing different decoding schemes to be compared under varying degrees of parallelism.
Taking serial SGRAND as the reference baseline, the acceleration factor $\eta_{\text{acc}}^{(\cdot)}$ is defined in \eqref{eq:eta}.
\begin{align}
    \Phi_{\text{tot}}^{(\cdot)} &= \Phi_{\text{heap}}^{(\cdot)} + \Phi_{\text{par}}^{(\cdot)}, \label{eq:Phi_tot}\\
    \Phi_{\text{eff}}^{(\cdot)} &= \Phi_{\text{heap}}^{(\cdot)} + \Phi_{\text{par}}^{(\cdot)}/P, \label{eq:Phi_eff}\\
    \eta_{\text{acc}}^{(\cdot)} &= \frac{\Phi_{\text{tot}}^{(\text{SG})}}{\Phi_{\text{eff}}^{(\cdot)}} = \frac{\Phi_{\text{tot}}^{(\text{SG})}}{\Phi_{\text{heap}}^{(\cdot)} + \Phi_{\text{par}}^{(\cdot)}/P}. \label{eq:eta}
\end{align}

Table~\ref{tab:grand_complexity2} reports the instantiated complexity results of different decoding schemes for the BCH$(127,106)$ code at $T=10^6$ and $E_b/N_0 = 5$~dB. We use RS-ORBGRAND as the representative method of ORB-type GRAND \cite{wan2024approaching}. In Table~\ref{tab:grand_complexity2}, the numbers in parentheses (e.g., ``Real Addition (8)'') denote the relative cost of each operation in XOR-equivalent units, and the table entries correspond to the average number of occurrences of each operation obtained via Monte Carlo simulations. The reported values are obtained by explicitly counting the number of operations during Monte Carlo decoding simulations, following the complexity model summarized in Table~\ref{tab:grand_complexity}, with the heap-related complexity $\Phi_{\text{heap}}^{(\cdot)}$ highlighted in bold italic.

By examining the complexity of parallel SGRAND under different parallelism degrees, it is observed that increasing the parallelism generally leads to a larger average number of tested EPs, thereby increasing the total computational complexity. At sufficiently high parallelism levels, heap sorting and tree-based operations become the dominant bottleneck. Consequently, an appropriate parallelism degree exists. For the considered setting, $P=32$ minimizes the effective computational complexity of parallel SGRAND as defined in \eqref{eq:Phi_eff}.

For ORBGRAND and the hybrid enhanced ORBGRAND algorithm, the same parallelism degree $P=32$ is adopted to ensure a fair comparison. Compared with ORBGRAND, the hybrid enhanced scheme introduces only a modest increase in heap-related operations, resulting in a limited rise in overall complexity. As a result, heap-related operations do not become the dominant bottleneck, while ML decoding performance is preserved.

\begin{figure}[tbp]
    \centering
     \includegraphics[width=0.46\textwidth]{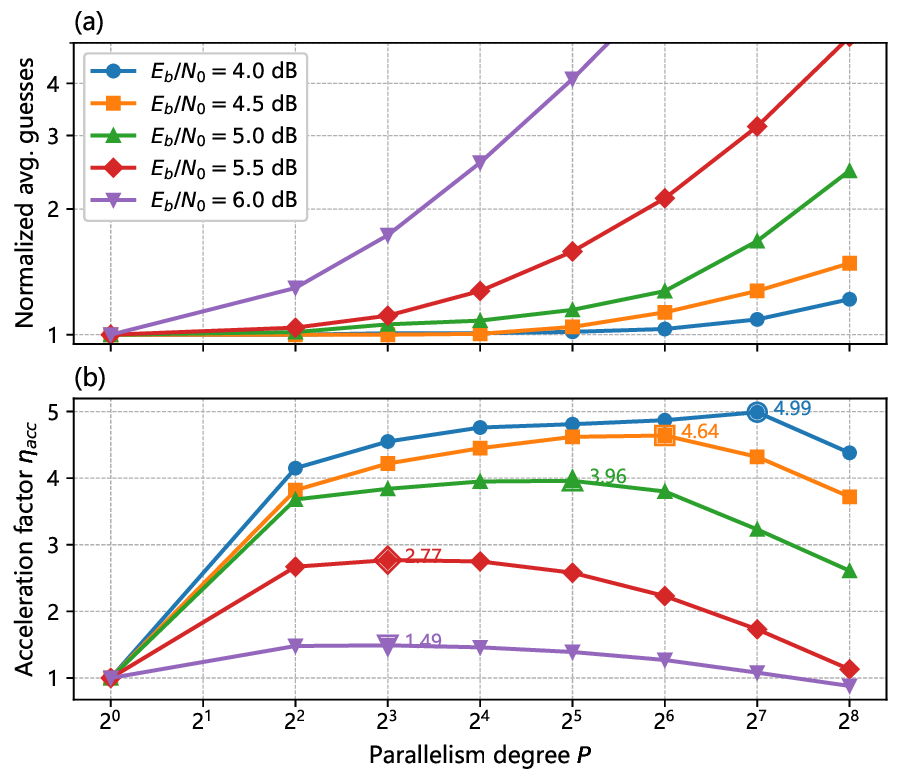}
    \caption{Performance comparison of parallel SGRAND with different degrees of parallelism: (a) Normalized average number of guesses (with $P=1$ as baseline); (b) Acceleration factor $\eta_{\text{acc}}$.}
    \label{fig:toy_model}
\end{figure}

Figure~\ref{fig:toy_model} provides an intuitive illustration of how the degree of parallelism influences the behavior of parallel SGRAND. 
Specifically, Fig.~\ref{fig:toy_model}(a) shows the normalized average number of queries, defined as the ratio between the average number of queries under parallelism degree $P$ and that for the serial case ($P=1$). Fig.~\ref{fig:toy_model}(b) depicts the corresponding acceleration factor $\eta_{\text{acc}}$. As parallelism increases, more EPs are tested concurrently in each decoding round, which jointly affects the total number of tested EPs and the achievable speedup.


Moreover, the optimal degree of parallelism is not fixed and varies with the operating SNR. At lower $E_b/N_0$, decoding typically requires a larger number of queries, and a higher degree of parallelism is therefore preferred to reduce the effective decoding latency.

\begin{figure}[tbp]
    \centering
    \includegraphics[width=0.48\textwidth]{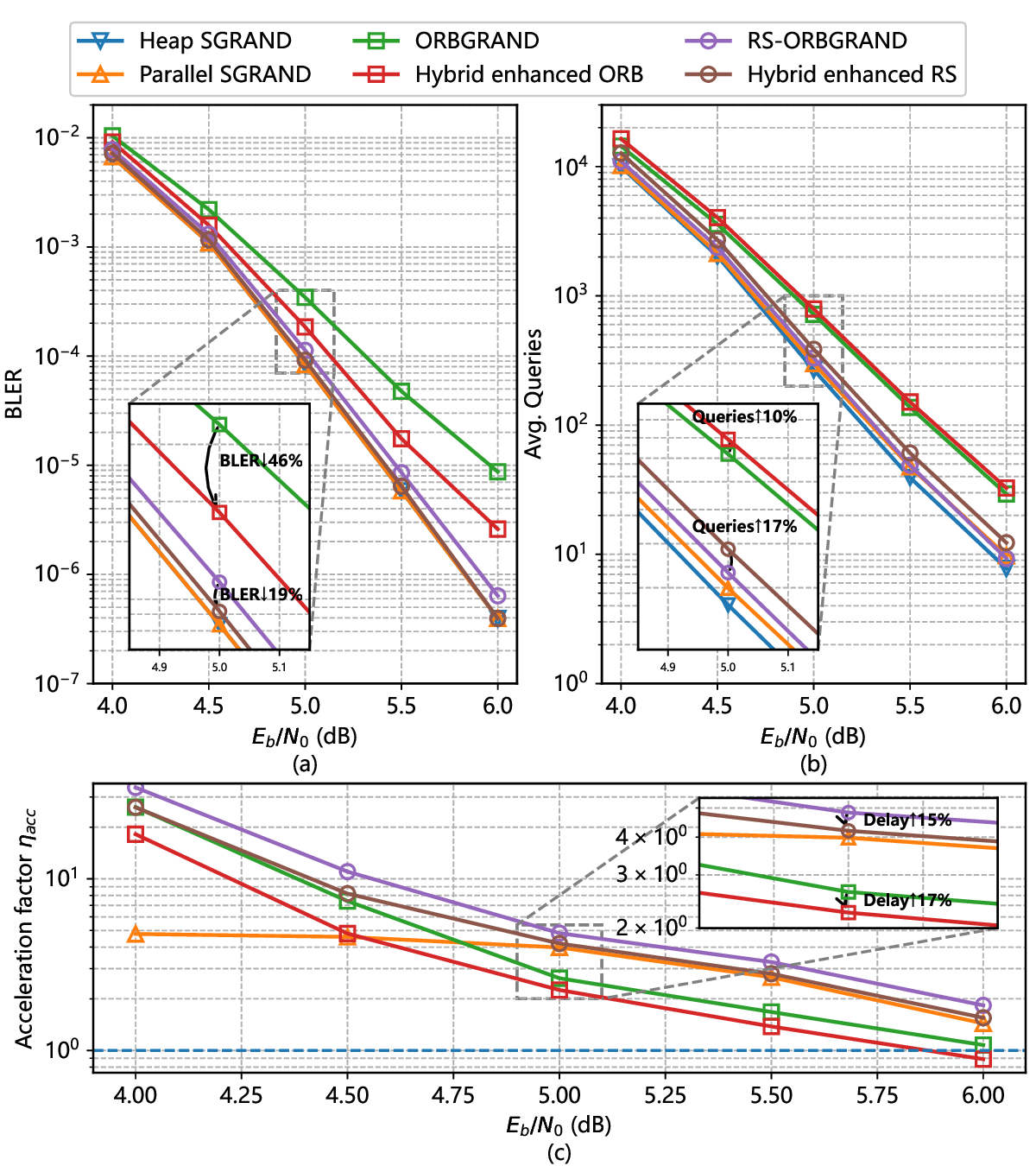}
    \caption{Performance evaluation of hybrid enhanced ORBGRAND: (a) BLER; (b) Average query number; (c) Acceleration factor $\eta_{\text{acc}}$.}
    \label{fig:3}
\end{figure}

Building upon the above observations, we next evaluate the performance of the proposed decoding framework. We fix the size of $\tilde{\mathcal{E}}_{\text{ORB}}$ to $T = 10^6$ and consider $E_b/N_0 \in \{4.0, 4.5, 5.0, 5.5, 6.0\}$~dB. The corresponding degrees of parallelism are set to $P \in \{128, 64, 32, 16, 8\}$, respectively. The results are shown in Fig.~\ref{fig:3}.

Figures~\ref{fig:3}(a) and (b) present the BLER and the average number of tested EPs as functions of $E_b/N_0$. As expected, serial and parallel SGRAND exhibit identical decoding performance, while the parallel implementation incurs a modest increase in the number of tested EPs due to its parallel search strategy. To assess the effectiveness of the proposed hybrid enhanced ORBGRAND, we compare it against two baseline ORB-type algorithms: ORBGRAND~\cite{duffy2022ordered} and RS-ORBGRAND~\cite{wan2024approaching}. Although these baseline algorithms exhibit different initial performance characteristics, both benefit substantially from the incorporation of the parallel SGRAND stage, which enhances their decoding performance toward ML optimality.

Unless the maximum query limit is set to $T = 2^N$, hybrid enhanced ORBGRAND cannot strictly guarantee ML decoding. Nevertheless, its performance improvement over standalone ORB-type decoding remains significant. For instance, at $E_b/N_0 = 5$~dB, the hybrid enhanced ORBGRAND achieves a 46\% reduction in BLER with only a 10\% increase in the number of tested EPs compared with ORBGRAND. When RS-ORBGRAND is used as the initial stage, the resulting hybrid scheme attains a BLER that closely approaches that of SGRAND with a 17\% increase in query number, further demonstrating the effectiveness of the proposed enhancement strategy.

Figure~\ref{fig:3}(c) presents the effective computational complexity analysis, with serial SGRAND used as the reference baseline. The hybrid enhanced ORBGRAND algorithms achieve substantial complexity reductions; for example, at $E_b/N_0 = 5$~dB, it achieves speedups of $2.63\times$ and $4.82\times$ when generating $\mathcal{E}_{\text{ORB}}$ using ORBGRAND and RS-ORBGRAND, respectively., respectively. Although the parallel SGRAND stage introduces additional overhead due to heap operations and the increased number of tested EPs, resulting in an approximately 15\% increase in effective computational complexity compared with ORB-type GRAND alone, the hybrid approach still achieves lower effective complexity than directly applying parallel SGRAND.

\begin{table*}[htbp]
\centering
\caption{Ablation study of acceleration techniques at a representative operating point ($E_b/N_0 = 5$~dB).
$\Phi_{\mathrm{eff}}$ reflects the effective computational complexity under parallel execution ($P=32$).}
\renewcommand{\arraystretch}{1.4}
\label{tab:ablation_flat}
\setlength{\tabcolsep}{4pt}
\begin{tabular}{l|cccc|cc}
\hline
Method 
& $\Phi_{\text{eff}}$ 
& $\Phi_{\text{tot}}$ 
& Avg. Queries
& $\Phi_{\text{eff}}$/Queries 
& $\eta$ (vs. Serial) 
& $\eta$ (vs. Basic Parallel) \\ 
\hline
Serial SGRAND
& -- 
& $1.59\times 10^5$
& $287$
& -- 
& $1.00\times$ 
& -- \\

Basic Parallel SGRAND (Sec.~\ref{sec:para_SGRAND})
& $4.91\times 10^4$
& $2.00\times 10^5$
& $380$
& $129$
& $3.23\times$
& $1.00\times$ \\

With Pruning (Sec.~\ref{subsec:5-1})
& $4.46\times 10^4$
& $1.82\times 10^5\;(\downarrow\!9.2\%)$
& $341\;(\downarrow\!10.3\%)$
& $131$
& $3.56\times$
& $1.10\times$ \\

With Tree Acceleration (Sec.~\ref{subsec:5-2})
& $4.53\times 10^4$
& $7.96\times 10^4\;(\downarrow\!60.3\%)$
& $380$
& $119\;(\downarrow\!7.7\%)$
& $3.50\times$
& $1.08\times$ \\

With Early Termination (Sec.~\ref{subsec:5-3})
& $4.70\times 10^4$
& $1.88\times 10^5\;(\downarrow\!6.2\%)$
& $353\;(\downarrow\!7.1\%)$
& $133$
& $3.38\times$
& $1.05\times$ \\

Parallel SGRAND (Alg.~\ref{alg:extended_parallel_SGRAND})
& $\mathbf{3.99\times 10^4}$
& $\mathbf{6.85\times 10^4\;(\downarrow\!65.8\%)}$
& $\mathbf{318\;(\downarrow\!16.3\%)}$
& $\mathbf{125\;(\downarrow\!3.0\%)}$
& $\mathbf{3.98\times}$
& $\mathbf{1.23\times}$ \\
\hline
\end{tabular}
\end{table*}

\subsection{Ablation Study of Acceleration Techniques}\label{subsec:6-3}

Table~\ref{tab:ablation_flat} reports the ablation results of the acceleration techniques introduced in Section~\ref{sec:extend_SGRAND} and integrated in Algorithm~\ref{alg:extended_parallel_SGRAND}. 
All results are obtained at a representative operating point of $E_b/N_0 = 5$~dB with a fixed parallelism factor $P=32$.

Taking the basic parallel SGRAND as the ablation baseline, pruning primarily reduces the number of tested EPs, yielding a $10.3\%$ reduction in the average queries. 
Early termination also suppresses unnecessary searches, further reducing the number of tests by 7.1\%. Using both approaches together reduces the number of queries by 16.3\%.
Both operations slightly increase the per-guess cost, as reflected by a modest rise in $\Phi_{\text{eff}}/\text{guess}$, due to the additional control and decision logic introduced at each evaluation. Nevertheless, the overall effective complexity $\Phi_{\text{eff}}$ is still reduced, since the decrease in the total number of tested EPs dominates.

Tree-based acceleration, on the other hand, directly targets the cost of each EP evaluation. 
By reusing reliability and parity computations in a structured manner, it reduces the algorithm-level workload by more than $60\%$, and correspondingly lowers the effective complexity per guess despite parallel amortization. 
When combined with pruning and early termination, the resulting parallel SGRAND benefits simultaneously from fewer tested EPs and a lower per-guess cost, achieving a cumulative reduction in effective complexity.

Overall, basic parallel SGRAND achieves a $3.23\times$ reduction in effective computational complexity compared with serial SGRAND through parallel execution alone. 
By integrating pruning, tree-based acceleration, and early termination, the proposed parallel SGRAND further improves this gain to $3.98\times$ while preserving ML optimality, confirming that the three techniques provide complementary and mutually reinforcing benefits.



\subsection{Generality Across Code Rates and Codebooks}\label{subsec:6-4}

\begin{table*}[htbp]
    \centering
    \caption{Performance Comparison of Algorithms under BCH(127,113). Three lowest BLER are highlighted in red, from darkest to lightest.}
    \label{tab:performance_comparison}
    \resizebox{\textwidth}{!}{
    \renewcommand{\arraystretch}{1.3}
    \begin{tabular}{lcccccccccccc}
        \toprule
        \multirow{2}{*}{\textbf{Algorithm}} 
        & \multicolumn{4}{c}{\textbf{BLER}} 
        & \multicolumn{8}{c}{\textbf{Average Queries (Avg / Ratio)}} \\
        \cmidrule(lr){2-5} \cmidrule(lr){6-13}
        & \textbf{@4 dB} & \textbf{@5 dB} & \textbf{@6 dB} & \textbf{@7 dB} 
        & \multicolumn{2}{c}{\textbf{@4 dB}} & \multicolumn{2}{c}{\textbf{@5 dB}} & \multicolumn{2}{c}{\textbf{@6 dB}} & \multicolumn{2}{c}{\textbf{@7 dB}} \\
        \midrule
        SGRAND \cite{solomon2020soft}          
        & \cellcolor{red!40}4.74\text{e}-2 & \cellcolor{red!40}2.37\text{e}-3 & \cellcolor{red!40}3.62\text{e}-5 & \cellcolor{red!40}2.70\text{e}-7 
        & 8.51\text{e}+2 & 1.00$\times$ 
        & 5.85\text{e}+1 & 1.00$\times$
        & 3.93\text{e}+0 & 1.00$\times$
        & 1.33\text{e}+0 & 1.00$\times$ \\
        ORBGRAND \cite{duffy2022ordered}        
        & 5.86\text{e}-2 & 4.72\text{e}-3 & 1.90\text{e}-4 & 4.78\text{e}-6 
        & 1.03\text{e}+3 & 1.21$\times$ 
        & 1.01\text{e}+2 & 1.73$\times$
        & 7.32\text{e}+0 & 1.86$\times$
        & 1.48\text{e}+0 & 1.11$\times$ \\
        Hybrid GRAND (ORB)      
        & \cellcolor{red!25}4.75\text{e}-2 & \cellcolor{red!25}2.39\text{e}-3 & \cellcolor{red!10}3.81\text{e}-5 & \cellcolor{red!10}3.20\text{e}-7 
        & 1.24\text{e}+3 & 1.46$\times$
        & 1.14\text{e}+2 & 1.95$\times$
        & 7.58\text{e}+0 & 1.93$\times$
        & 1.48\text{e}+0 & 1.11$\times$ \\
        RS-ORBGRAND \cite{wan2024approaching}     
        & 5.10\text{e}-2 & 2.83\text{e}-3 & 5.21\text{e}-5 & 5.90\text{e}-7 
        & 9.30\text{e}+2 & 1.09$\times$ 
        & 6.82\text{e}+1 & 1.17$\times$
        & 4.50\text{e}+0 & 1.15$\times$
        & 1.37\text{e}+0 & 1.03$\times$ \\
        Hybrid GRAND (RS)   
        & \cellcolor{red!40}4.74\text{e}-2 & \cellcolor{red!40}2.37\text{e}-3 & \cellcolor{red!25}3.68\text{e}-5 & \cellcolor{red!25}2.80\text{e}-7 
        & 1.09\text{e}+3 & 1.28$\times$ 
        & 7.74\text{e}+1 & 1.32$\times$
        & 4.67\text{e}+0 & 1.19$\times$
        & 1.37\text{e}+0 & 1.03$\times$ \\
        List GRAND \cite{abbas2022list}       
        & 5.10\text{e}-2 & 2.72\text{e}-3 & 5.10\text{e}-5 & 4.30\text{e}-7 
        & 8.42\text{e}+3 & 9.89$\times$ & 8.01\text{e}+2 & 13.7$\times$ & 6.00\text{e}+1 & 15.2$\times$ & 6.32\text{e}+0 & 4.75$\times$ \\
        MLD Lower Bound
        & 4.51\text{e}-2 & 2.19\text{e}-3 & 3.59\text{e}-5 & 2.70\text{e}-7 & -- & -- & -- & -- & -- & -- & -- & --  \\
        \bottomrule
    \end{tabular}
    }
\end{table*}

To further validate the effectiveness of the hybrid enhanced ORBGRAND, we increase the code rate to BCH$(127,113)$ and set the maximum query limit to $T=5\times10^{4}$, under which SGRAND is almost guaranteed to identify a valid EP. Under this setting, the decoding performance of SGRAND closely approaches that of ML decoding, thereby providing a practical reference for near-ML performance.

Table~\ref{tab:performance_comparison} summarizes the resulting BLER and the average number of queries. The reported ML lower bound is obtained under an optimistic assumption that decoding is declared successful even when no valid codeword is found during the SGRAND search. As expected, this bound is nearly indistinguishable from the performance of SGRAND when a sufficiently large number of queries is allowed. The ``Ratio'' column reports the average query number of each algorithm normalized by that of serial SGRAND.

The experimental results further show that, with a sufficiently large test budget, the proposed hybrid enhanced ORBGRAND algorithm can achieve near-ML performance when either ORBGRAND or RS-ORBGRAND is employed as the first-stage decoder, with only a modest increase in the average number of tests.

For comparison, we also include the experimental results of list GRAND \cite{abbas2022list}, which outputs multiple candidate valid EPs and selects the one with the highest reliability. The results indicate that list GRAND requires a substantially larger number of average tests and still fails to approach ML decoding performance, thereby highlighting the performance–complexity advantage of the proposed hybrid enhanced ORBGRAND algorithm.
\begin{figure}[tbp]
    \centering
    \includegraphics[width=0.48\textwidth]{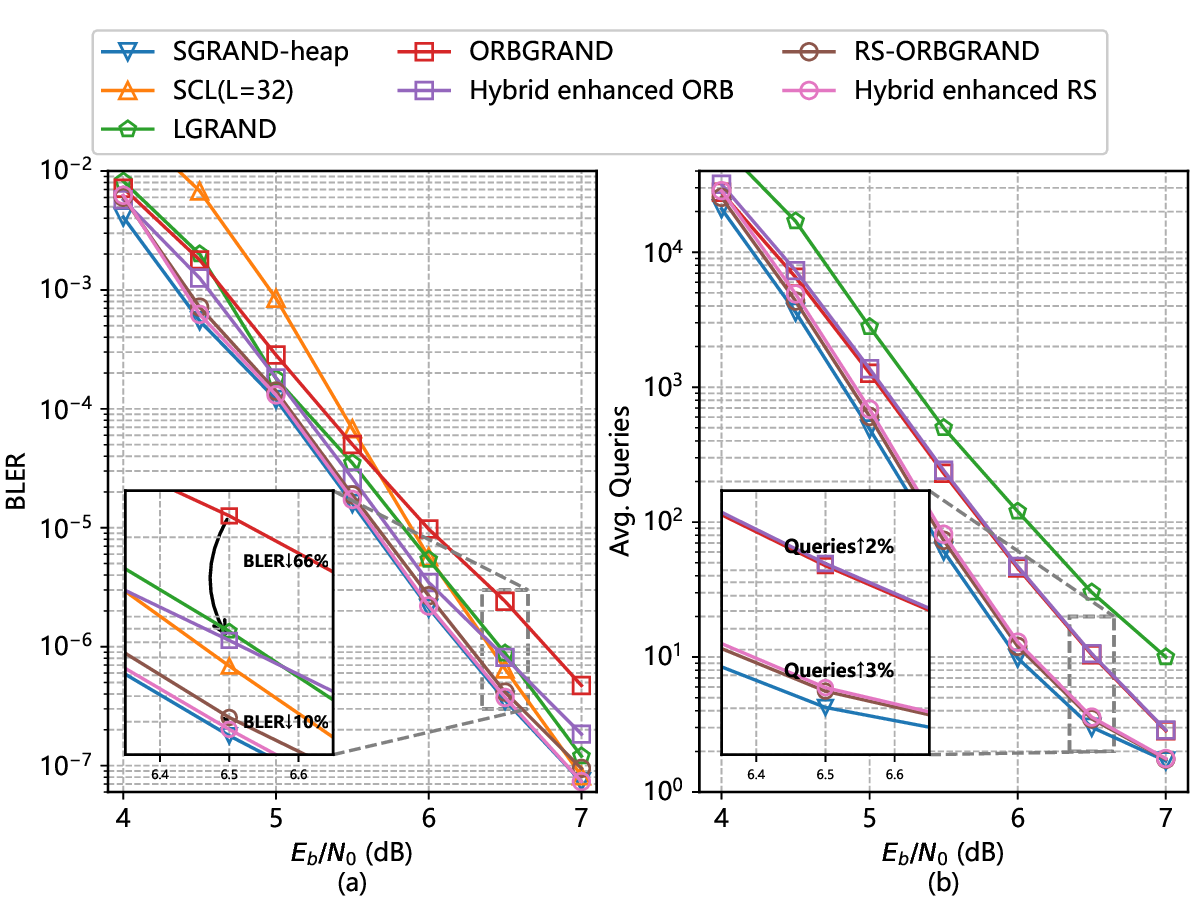}
    \caption{Comparison of decoding performance and average complexity of different GRAND varients for CA-polar code (128, 105+11): (a) BLER; (b) Average query number.}
    \label{fig:polar}
\end{figure}

We further evaluate the proposed hybrid enhanced ORBGRAND framework on the 5G NR CA-polar code $(128,105+11)$  and compare it with the CA-SCL decoder \cite{arikan2009channel,niu2012crc}. The maximum query budget is set to $T=3\times10^{6}$. For list GRAND, the results from~\cite{abbas2022list} with parameters $(LM_{\text{max}}=96,\ HW_{\text{max}}=8,\ \delta=20)$ are reported.

Fig.~\ref{fig:polar} presents the BLER and average number of queries versus $E_b/N_0$. As shown in Fig.~\ref{fig:polar}(a), ORBGRAND exhibits a clear performance gap to the near-ML reference. With hybrid enhancement, the BLER of ORBGRAND is reduced by approximately $66\%$ in the highlighted high-SNR region. A similar improvement is observed for RS-ORBGRAND, where hybrid enhancement achieves about a $10\%$ BLER reduction and further narrows the gap to ML decoding.

These performance gains are achieved with only a modest increase in query complexity. As shown in Fig.~\ref{fig:polar}(b), hybrid enhanced ORBGRAND increases the average number of queries by about $2\%$, while the increase for hybrid enhanced RS-ORBGRAND is limited to approximately $3\%$. In contrast, list GRAND requires substantially more queries across the entire SNR range and still fails to approach near-ML performance.

Overall, the results on the CA-polar code confirm that the proposed hybrid enhancement significantly improves ORB-type GRAND decoding with minimal additional query overhead, yielding a more favorable performance--complexity tradeoff than list GRAND.

\section{Conclusion}\label{sec:conclusion}



This paper presents a theoretical investigation of the parallel implementation of SGRAND while preserving its ML decoding property. By exploiting the EP tree structure, we further propose a hybrid decoding approach that integrates ORBGRAND with SGRAND, enabling ORBGRAND to approach ML decoding performance. Owing to the parallel design of SGRAND, the resulting hybrid algorithm can also be implemented efficiently under parallel execution.

Through decoding performance simulations and complexity analysis, the proposed parallel algorithms are shown to be both effective and practical. In particular, the hybrid enhanced GRAND scheme achieves substantial performance improvements while requiring only about a $10\%$ increase in the average number of queries, thereby significantly narrowing the gap to ML decoding. Nevertheless, fully realizing the potential of parallelism requires dedicated hardware implementations. Developing such architectures to enable real-time ML decoding for URLLC applications constitutes an important direction for future research.

\bibliographystyle{IEEEtran}
\bibliography{Ref_abbreviation}

@article{shirvanimoghaddam2018short,
  title={Short block-length codes for ultra-reliable low latency communications},
  author={Shirvanimoghaddam, Mahyar and Mohammadi, Mohammad Sadegh and Abbas, Rana and Minja, Aleksandar and Yue, Chentao and Matuz, Balazs and Han, Guojun and Lin, Zihuai and Liu, Wanchun and Li, Yonghui and others},
  journal={IEEE Commun. Mag.},
  volume={57},
  number={2},
  pages={130--137},
  year={2018},
  publisher={IEEE}
}

@article{yue2023efficient,
  title={Efficient decoders for short block length codes in {6G} {URLLC}},
  author={Yue, Chentao and Miloslavskaya, Vera and Shirvanimoghaddam, Mahyar and Vucetic, Branka and Li, Yonghui},
  journal={IEEE Commun. Mag.},
  volume={61},
  number={4},
  pages={84--90},
  year={2023},
  publisher={IEEE}
}

@article{wang2023road,
  title={On the road to {6G}: Visions, requirements, key technologies, and testbeds},
  author={Wang, Cheng-Xiang and You, Xiaohu and Gao, Xiqi and Zhu, Xiuming and Li, Zixin and Zhang, Chuan and Wang, Haiming and Huang, Yongming and Chen, Yunfei and Haas, Harald and others},
  journal={IEEE Commun. Surveys Tuts.},
  volume={25},
  number={2},
  pages={905--974},
  year={2023},
  publisher={IEEE}
}

@article{gautam2025analysis,
  author = {Gautam, A. and Thakur, P. and Singh, G.},
  title = {Analysis of Universal Decoding Techniques for {6G} Ultra-Reliable and Low-Latency Communication Scenario},
  journal = {Future Internet},
  volume = {17},
  number = {4},
  pages = {181},
  year = {2025}
}

@book{Lin2004ErrorCC,
    author ={Shu Lin and Daniel J. Costello},
    title = {Error Control Coding: Fundamentals and Applications},
    publisher = {Englewood Cliffs, NJ, USA: Prentice Hall},
    year = {2004}
}

@article{bose1960class,
  title={On a class of error correcting binary group codes},
  author={Bose, Raj Chandra and Ray-Chaudhuri, Dwijendra K},
  journal={Inf. Control},
  volume={3},
  number={1},
  pages={68--79},
  year={1960},
  publisher={Elsevier}
}

@article{arikan2009channel,
  title={Channel polarization: A method for constructing capacity-achieving codes for symmetric binary-input memoryless channels},
  author={Arikan, Erdal},
  journal={IEEE Trans. Inf. Theory},
  volume={55},
  number={7},
  pages={3051--3073},
  year={2009},
  publisher={IEEE}
}

@article{niu2012crc,
  title={{CRC}-aided decoding of polar codes},
  author={Niu, Kai and Chen, Kai},
  journal={IEEE Commun. Lett.},
  volume={16},
  number={10},
  pages={1668--1671},
  year={2012},
  publisher={IEEE}
}

@article{luby2002improved,
  title={Improved low-density parity-check codes using irregular graphs},
  author={Luby, Michael G and Mitzenmacher, Michael and Shokrollahi, Mohammad Amin and Spielman, Daniel A},
  journal={IEEE Trans. Inf. Theory},
  volume={47},
  number={2},
  pages={585--598},
  year={2002},
  publisher={IEEE}
}

@article{shao2019survey,
  title={{Survey of turbo, LDPC, and polar decoder ASIC implementations}},
  author={Shao, Shuai and Hailes, Peter and Wang, Tsang-Yi and Wu, Jwo-Yuh and Maunder, Robert G and Al-Hashimi, Bashir M and Hanzo, Lajos},
  journal={IEEE Commun. Surveys Tuts.},
  volume={21},
  number={3},
  pages={2309--2333},
  year={2019},
  publisher={IEEE}
}

@book{abbas2023guessing,
  title={Guessing random additive noise decoding: A hardware perspective},
  author={Abbas, Syed Mohsin and Jalaleddine, Marwan and Gross, Warren J},
  year={2023},
  publisher={Springer Nature}
}

@article{leroux2012semi,
  title={A semi-parallel successive-cancellation decoder for polar codes},
  author={Leroux, Camille and Raymond, Alexandre J and Sarkis, Gabi and Gross, Warren J},
  journal={IEEE Trans. Signal Process.},
  volume={61},
  number={2},
  pages={289--299},
  year={2012},
  publisher={IEEE}
}

@article{tarver2021gpu,
  title={{GPU}-based, {LDPC} decoding for {5G} and beyond},
  author={Tarver, Chance and Tonnemacher, Matthew and Chen, Hao and Zhang, Jianzhong and Cavallaro, Joseph R},
  journal={IEEE Open J. Circuits Syst.},
  volume={2},
  pages={278--290},
  year={2021},
  publisher={IEEE}
}

@article{fossorier2002soft,
  title={Soft-decision decoding of linear block codes based on ordered statistics},
  author={Fossorier, Marc PC and Lin, Shu},
  journal={IEEE Trans. Inf. Theory},
  volume={41},
  number={5},
  pages={1379--1396},
  year={2002},
  publisher={IEEE}
}

@inproceedings{afisiadis2014low,
  title={A low-complexity improved successive cancellation decoder for polar codes},
  author={Afisiadis, Orion and Balatsoukas-Stimming, Alexios and Burg, Andreas},
  booktitle={Proc. 48th Asilomar Conf. Signals, Syst. Comput. (ASILOMAR)},
  pages={2116--2120},
  year={2014},
  organization={IEEE}
}

@inproceedings{zheng2024locally,
  title={Locally constrained guessing codeword decoding of short block codes},
  author={Zheng, Xiangping and Ma, Xiao},
  booktitle={Proc. IEEE Inf. Theory Workshop (ITW)},
  pages={442--447},
  year={2024}
}

@article{zheng2024universal,
  title={A universal list decoding algorithm with application to decoding of polar codes},
  author={Zheng, Xiangping and Ma, Xiao},
  journal={IEEE Trans. Inf. Theory},
  volume={71},
  number={2},
  pages={975--995},
  year={2024},
  publisher={IEEE}
}

@article{wang2025two,
  title={A two-stage soft-decision decoding algorithm for {BCH} codes},
  author={Wang, Qianfan and Wang, Yiwen and Liang, Jifan and Song, Linqi and Ma, Xiao},
  journal={IEEE Trans. Commun.},
  volume={73},
  number={12},
  pages={12979--12992},
  year={2025},
  publisher={IEEE}
}

@article{wang2025guessing,
  title={Guessing Decoding of Short Blocklength Codes},
  author={Wang, Qianfan and Liang, Jifan and Yuan, Peihong and Duffy, Ken R and M{\'e}dard, Muriel and Ma, Xiao},
  journal={arXiv preprint arXiv:2511.12108},
  year={2025}
}

@article{dorsch1974decoding,
  title={A decoding algorithm for binary block codes and {J}-ary output channels (corresp.)},
  author={Dorsch, B},
  journal={IEEE Trans. Inf. Theory},
  volume={20},
  number={3},
  pages={391--394},
  year={1974},
  publisher={IEEE}
}

@article{yue2021probability,
  title={Probability-based ordered-statistics decoding for short block codes},
  author={Yue, Chentao and Shirvanimoghaddam, Mahyar and Park, Giyoon and Park, Ok-Sun and Vucetic, Branka and Li, Yonghui},
  journal={IEEE Commun. Lett.},
  volume={25},
  number={6},
  pages={1791--1795},
  year={2021},
  publisher={IEEE}
}

@inproceedings{pillet2022classification,
  title={Classification of automorphisms for the decoding of polar codes},
  author={Pillet, Charles and Bioglio, Valerio and Land, Ingmar},
  booktitle={Proc. IEEE Int. Conf. Commun. (ICC)},
  pages={110--115},
  year={2022}
}

@article{liang2024random,
  title={A random coding approach to performance analysis of the ordered statistic decoding with local constraints},
  author={Liang, Jifan and Ma, Xiao},
  journal={arXiv preprint arXiv:2401.16709},
  year={2024}
}

@article{duffy2019capacity,
  title={Capacity-achieving guessing random additive noise decoding},
  author={Duffy, Ken R and Li, Jiange and M{\'e}dard, Muriel},
  journal={IEEE Trans. Inf. Theory},
  volume={65},
  number={7},
  pages={4023--4040},
  year={2019},
  publisher={IEEE}
}

@inproceedings{solomon2020soft,
  title={Soft maximum likelihood decoding using {GRAND}},
  author={Solomon, Amit and Duffy, Ken R and M{\'e}dard, Muriel},
  booktitle={Proc. IEEE Int. Conf. Commun. (ICC)},
  pages={1--6},
  year={2020}
}

@inproceedings{condo2021iLWO,
  title={High-performance low-complexity error pattern generation for {ORBGRAND} decoding},
  author={Condo, Carlo and Bioglio, Valerio and Land, Ingmar},
  booktitle={Proc. IEEE Globecom Workshops},
  pages={1--6},
  year={2021}
}

@article{duffy2022ordered,
  title={Ordered reliability bits guessing random additive noise decoding},
  author={Duffy, Ken R and An, Wei and M{\'e}dard, Muriel},
  journal={IEEE Trans. Signal Process.},
  volume={70},
  pages={4528--4542},
  year={2022},
  publisher={IEEE}
}

@article{abbas2022list,
  title={List-{GRAND}: A practical way to achieve maximum likelihood decoding},
  author={Abbas, Syed Mohsin and Jalaleddine, Marwan and Gross, Warren J},
  journal={IEEE Trans. Very Large Scale Integr. (VLSI) Syst.},
  volume={31},
  number={1},
  pages={43--54},
  year={2022},
  publisher={IEEE}
}

@article{liu2022orbgrand,
  title={{ORBGRAND} is almost capacity-achieving},
  author={Liu, Mengxiao and Wei, Yuejun and Chen, Zhenyuan and Zhang, Wenyi},
  journal={IEEE Trans. Inf. Theory},
  volume={69},
  number={5},
  pages={2830--2840},
  year={2022},
  publisher={IEEE}
}

@INPROCEEDINGS{wan2024approaching,
  author={Wan, Li and Zhang, Wenyi},
  booktitle={Proc. IEEE Int. Symp. on Inf. Theory}, 
  title={Approaching Maximum Likelihood Decoding Performance via Reshuffling {ORBGRAND}}, 
  year={2024},
  volume={},
  number={},
  pages={31-36}
}

@inproceedings{li2024orbgrand,
  title={{ORBGRAND}: Achievable Rate for General Bit Channels and Application in {BICM}},
  author={Li, Zhuang and Zhang, Wenyi},
  booktitle={Proc. IEEE Int. Symp. Pers., Indoor Mobile Radio Commun. (PIMRC)},
  pages={1--7},
  year={2024}
}

@INPROCEEDINGS{wan2025finetuning,
  author={Wan, Li and Yin, Huarui and Zhang, Wenyi},
  booktitle={Proc. IEEE/CIC ICCC Workshops}, 
  title={Fine-tuning {ORBGRAND} with Very Few Channel Soft Values}, 
  year={2025},
  volume={},
  number={},
  pages={1-6}
}

@inproceedings{yuan2023guessing,
  title={Guessing random additive noise decoding with quantized soft information},
  author={Yuan, Peihong and Duffy, Ken R and Gabhart, Evan P and M{\'e}dard, Muriel},
  booktitle={Proc. IEEE Globecom Workshops},
  pages={1698--1703},
  year={2023}
}

@inproceedings{abbas2020high,
  title={High-throughput {VLSI} architecture for {GRAND}},
  author={Abbas, Syed Mohsin and Tonnellier, Thibaud and Ercan, Furkan and Gross, Warren J},
  booktitle={Proc. IEEE Workshop Signal Process. Syst. (SiPS)},
  pages={1--6},
  year={2020}
}

@ARTICLE{9715725,
  author={Condo, Carlo},
  journal={IEEE Trans. Circuits Syst. I: Regul. Papers}, 
  title={A Fixed Latency {ORBGRAND} Decoder Architecture With {LUT}-Aided Error-Pattern Scheduling}, 
  year={2022},
  volume={69},
  number={5},
  pages={2203-2211}
}

@article{abbas2022high,
  title={High-throughput and energy-efficient {VLSI} architecture for ordered reliability bits {GRAND}},
  author={Abbas, Syed Mohsin and Tonnellier, Thibaud and Ercan, Furkan and Jalaleddine, Marwan and Gross, Warren J},
  journal={IEEE Trans. Very Large Scale Integr. (VLSI) Syst.},
  volume={30},
  number={6},
  pages={681--693},
  year={2022},
  publisher={IEEE}
}

@inproceedings{xiao2023low,
  title={A low-latency and area-efficient {ORBGRAND} decoder for polar codes},
  author={Xiao, Jiayu and Zhou, Yangcan and Song, Suwen and Wang, Zhongfeng},
  booktitle={Proc. 4th Inf. Commun. Technol. Conf. (ICTC)},
  pages={10--15},
  year={2023},
  organization={IEEE}
}

@article{ji2024efficient,
  title={Efficient {ORBGRAND} Implementation With Parallel Noise Sequence Generation},
  author={Ji, Chao and You, Xiaohu and Zhang, Chuan and Studer, Christoph},
  journal={IEEE Trans. Very Large Scale Integr. (VLSI) Syst.},
  volume={33},
  number={2},
  pages={435--448},
  year={2025},
  publisher={IEEE}
}

@article{abbas2025improved,
  title={Improved {Step-GRAND}: Low-Latency Soft-Input Guessing Random Additive Noise Decoding},
  author={Abbas, Syed Mohsin and Jalaleddine, Marwan and Tsui, Chi-Ying and Gross, Warren J},
  journal={IEEE Trans. Very Large Scale Integr. (VLSI) Syst.},
  volume={33},
  number={4},
  pages={1028--1041},
  year={2025},
  publisher={IEEE}
}

@inproceedings{riaz2023subpj,
  title={A sub-0.8 p{J}/b 16.3 {G}bps/mm$^2$ universal soft-detection decoder using {ORBGRAND} in 40nm {CMOS}},
  author={Riaz, Arslan and Yasar, Alperen and Ercan, Furkan and An, Wei and Ngo, Jonathan and Galligan, Kevin and Medard, Muriel and Duffy, Ken R and Yazicigil, Rabia Tugce},
  booktitle={Proc. IEEE Int. Solid-State Circuits Conf. (ISSCC)},
  pages={432--434},
  year={2023}
}

@article{riaz2025subpj,
  title={A sub-0.8-p{J}/bit universal soft-detection decoder using {ORBGRAND}},
  author={Riaz, Arslan and Yasar, Alperen and Ercan, Furkan and An, Wei and Ngo, Jonathan and Galligan, Kevin and M{\'e}dard, Muriel and Duffy, Ken R and Yazicigil, Rabia Tugce},
  journal={IEEE J. Solid-State Circuits},
  volume={60},
  number={7},
  pages={2645--2659},
  year={2024},
  publisher={IEEE}
}

@inproceedings{kizilates2025lowlatency,
  title={Low-latency modulation-and correlation-adaptive {ORBGRAND}-{AI} decoder},
  author={Kizilates, Zeynep Ece and Riaz, Arslan and Bali, Akshaya and Grundei, Moritz and Medard, Muriel and Duffy, Ken R and Yazicigil, Rabia Tugce},
  booktitle={Proc. IEEE Eur. Solid-State Electron. Res. Conf. (ESSERC)},
  pages={641--644},
  year={2025}
}

@article{taipale1991improvement,
  title={An improvement to generalized-minimum-distance decoding},
  author={Taipale, Dana John and Pursley, Michael B},
  journal={IEEE Trans. Inf. Theory},
  volume={37},
  number={1},
  pages={167--172},
  year={1991},
  publisher={IEEE}
}

@book{cormen2022introduction,
  title={Introduction to algorithms},
  author={Cormen, Thomas H and Leiserson, Charles E and Rivest, Ronald L and Stein, Clifford},
  year={2022},
  publisher={MIT press}
}

\begin{IEEEbiography}[{\includegraphics[width=1in,height=1.25in,clip,keepaspectratio]{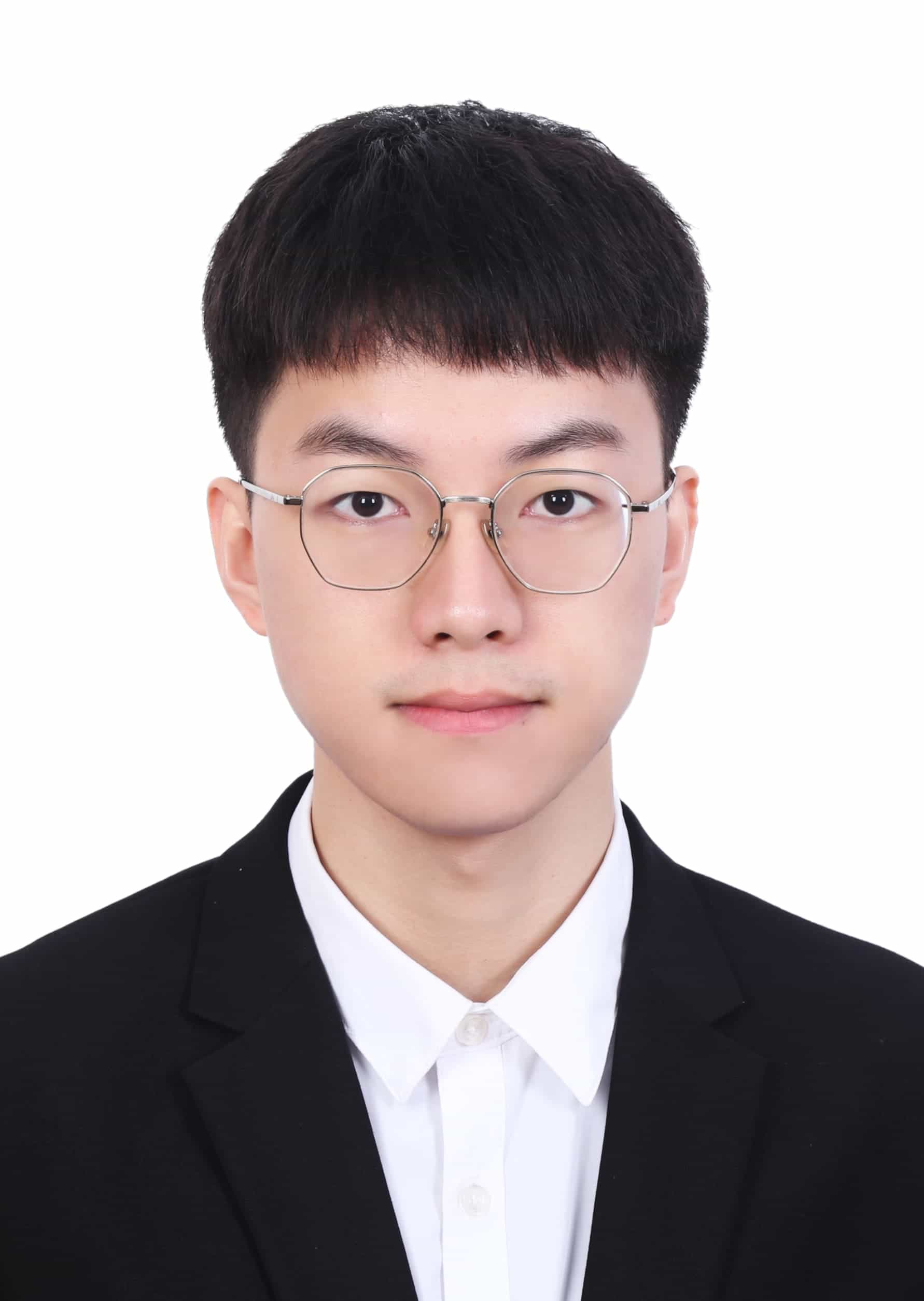}}]{Li Wan}
    received the B.S. degree in statistics from the University of Science and Technology of China (USTC), Hefei, China, in 2021, where he is currently pursuing the Ph.D. degree with the Department of Electronic Engineering and Information Science. His research interests lie in information theory and coding theory.
\end{IEEEbiography}

\begin{IEEEbiography}[{\includegraphics[width=1in,height=1.25in,clip,keepaspectratio]{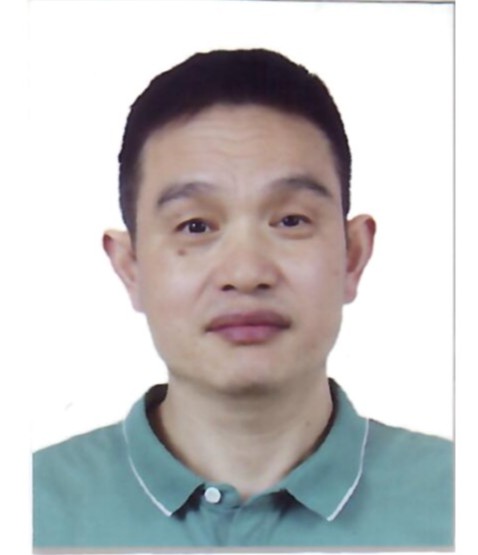}}]{Huarui Yin}
    received the B.S. degree in 1996 and the Ph.D. degree in 2006, both from the University of Science and Technology of China (USTC), Hefei, China. He is currently an Associate Professor with the Department of Electronic Engineering and Information Science, USTC. His research interests include wireless communications, signal processing, and communication theory.
\end{IEEEbiography}

\begin{IEEEbiography}[{\includegraphics[width=1in,height=1.25in,clip,keepaspectratio]{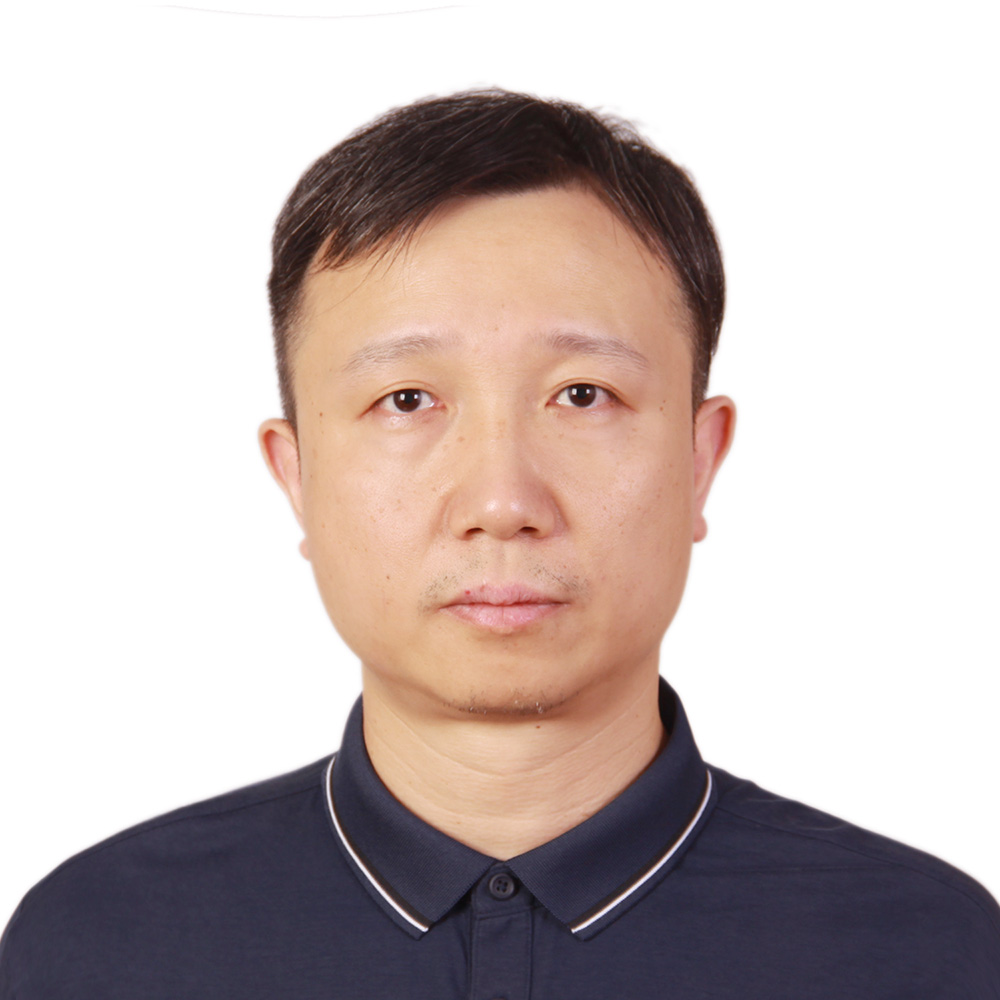}}]{Wenyi Zhang} (Senior Member, IEEE) received the bachelor's degree from the Department of Automation, Tsinghua University, in 2001, and the master's and Ph.D. degrees from the Department of Electrical Engineering, University of Notre Dame, in 2003 and 2006, respectively. Since January 2010, he has been a Faculty Member with the Department of Electronic Engineering and Information Science, University of Science and Technology of China, where he is currently a Professor. Prior to that, he was with the Communications Science Institute, University of Southern California, as a Post-Doctoral Research Associate, from September 2006 to April 2008, and Qualcomm Inc., Corporate Research and Development, from May 2008 to December 2009. His research interests lie at the intersection of communications, information theory, and statistical inference. He is an Editor of IEEE TRANSACTIONS ON INFORMATION THEORY and IEEE TRANSACTIONS ON COMMUNICATIONS. He is a Distinguished Lecturer of the IEEE Information Theory Society from 2026 to 2027.
\end{IEEEbiography}

\end{document}